\documentclass[runningheads]{llncs}
\usepackage[normalem]{ulem}
\usepackage[T1]{fontenc}
\usepackage{tabularx} 
\usepackage{booktabs}

\usepackage{multirow}
\usepackage{threeparttable}
\usepackage{float}   
\usepackage{siunitx} 
\usepackage{graphicx}
\usepackage{xcolor}
\newif\ifcomment
\commenttrue
\ifcomment
\newcommand{\Kobra}[1]{{\bf \textcolor{purple}{Kobra: #1}}}
\newcommand{\Raphael}[1]{{\bf \textcolor{blue}{Raphael: #1}}}
\newcommand{\Pooria}[1]{{\bf \textcolor{orange}{Pooria: #1}}}
\else
\newcommand{\Kobra}[1]{}
\newcommand{\Raphael}[1]{}
\newcommand{\Pooria}[1]{}
\fi
\newcounter{finding}

\newcolumntype{Y}{>{\centering\arraybackslash}X}

\usepackage{subcaption}

\begin{document}
\title{WildCode Revisited: A Comprehensive Empirical Study on the Security of LLM-Generated Code}

%

\author{
Kobra Khanmohammadi\inst{1}\orcidID{0009-0004-1414-2111} \and
Pooria Roy\inst{2}\orcidID{0009-0004-3990-9905} \and
Raphaël Khoury\inst{3}\orcidID{0000-0002-7625-3384} \and
Abdelwahab Hamou-Lhadj\inst{4}\orcidID{0000-0002-3319-5006} \and 
Wilfried Patrick Konan\inst{3} \and
Alexander Da Re\inst{1} \and
Nicholas Rebelo Melo\inst{1}\orcidID{0009-0005-8090-1523}
}

\authorrunning{K. Khanmohammadi et al.}

\institute{
Sheridan College, Ontario, Canada\\
\email{kobra.khanmohammadi@sheridancollege.ca, dareal@sheridancollege.ca, melnicho@sheridancollege.ca}
\and
School of Computing, Queen's University, Kingston, Canada\\
\email{pooria.roy@queensu.ca}
\and
Université du Québec en Outaouais (UQO), Canada\\
\email{raphael.khoury@uqo.ca, konk14@uqo.ca}
\and
Concordia University, Montreal, Canada\\
\email{wahab.hamou-lhadj@concordia.ca}
}

\maketitle              

\begin{abstract}
LLM models are increasingly used to generate code, but the quality and security of this code are often uncertain. Several recent studies have raised alarm bells, indicating that such AI-generated code may be particularly vulnerable to cyberattacks. However, most of these studies rely on code that is generated specifically for the study, which raises questions about the realism of such experiments. In this study, we perform a large-scale empirical analysis of real-life code generated by ChatGPT. We evaluate code generated by ChatGPT both with respect to correctness and security and delve into the intentions of users who request code from the model. We further performed an experiment to evaluate the effectiveness of common prompt engineering strategies using real-life prompts. Our study supports earlier research that employed synthetic queries and produced proof that LLM-generated code is frequently insufficient in terms of security. Additionally, we observe that users don't ask many questions about the security characteristics of the code they ask LLMs to provide.

\end{abstract}
\section{Introduction}
In the span of only a few years, LLMs went from an emerging technology to an everyday tool, widely used by both tech-savvy programmers and ordinary users alike. Of particular interest is the use of LLMs to generate code. Recent surveys indicate that most developers rely on LLMs to generate code \cite{Etsenake2024}, a trend so pronounced that it is even blamed for a slowdown in the hiring of programmers.   

This rapid change in the practice of computer programming took place with little consideration of the security of the code being produced. The preliminary findings on the degree of security of the code produced by LLMs are alarming \cite{khoury2023secure,fu2025security,tihanyi2023new}. Such studies usually find that LLMs produce code that falls below even modest expectations of security and may require extensive modifications before it can be run safely in an untrusted context. This is especially the case if the programmer does not explicitly request that the code contains security checks or that it be resistant to specific categories of attack \cite{khoury2023secure}.

Most initial studies on the topic proceeded by asking an LLM to generate a series of programs, often guided by specific scenarios, and analyzing the resulting code, either manually or using an automated tool \cite{zeng2025inducing,he2023large,li2024cosec,bai2024apilot,nazzal2024promsec}. While such studies provide useful insights, there is a threat to validity because the scenarios chosen may not be representative of the actual interactions that programmers have with LLMs. 

In this paper, we present the first empirical study on the security of code generated by ChatGPT, one of the most widely used LLMs. Our analysis is based on data extracted from WildChat \cite{zhao2024wildchat}, a publicly available dataset containing more than one million real-world conversations with ChatGPT,  from which we extract all conversations that include the code generated by the model. Unlike previous work that simulates user interactions by querying LLMs with synthetic prompts, our study leverages authentic user–ChatGPT interactions. This enables us to examine not only the security of the generated code, but also what users intend to ask, how they follow up on ChatGPT responses, and how they react when encountering buggy or insecure code. We further investigate the prevalence of hallucinations and the presence of confidential data in the code.  Furthermore, we perform an investigation of the effectiveness of common prompt engineering strategies, again using edited real-life prompts from the WildChat dataset rather than synthetic prompts, as is common in multiple studies on this topic. 

In addition, we provide a curated set of annotated conversations and corresponding code snippets in which ChatGPT produced buggy or insecure responses. This dataset, available on our HuggingFace repository\footnote{https://huggingface.co/datasets/regularpooria/wildcode} and GitHub repository\footnote{https://github.com/regularpooria/wildcode}, contains the full list of annotated conversations and related code samples used in this study, as well as the rules used to extract patterns from the conversations, allowing reproducibility and facilitating further research in this area. The dataset includes syntactically correct code for Python, JavaScript, C/C++, Java, PHP, and C\#, as well as unchecked or potentially erroneous code for other languages.

This paper is an extended version of a paper presented at the 18th International Symposium on Foundations \& Practice of Security in November 2025 \cite{original}. The major extensions include the following:
\begin{itemize}
    \item A more thorough review of the relevant academic literature (Sect. \ref{sec:litt};
    \item An analysis of confidential information (keys and other secrets) present in the prompts (Sect. \ref{sec:secrets});
    \item A re-prompting experiment that examines the effectiveness of several widely used prompt engineering strategies on the real-life prompts from the Wildchat dataset (Sec. \ref{sec:reprompt});
    \item A test of hallucinations in more recent LLMs to determine whether architectural modifications and updated data have reduced library hallucinations (Sec. \ref{sec:hallucinations}).
    \item An analysis of the reasons why different models refused to generate code as requested (Sect. \ref{sec:Refusal}). 
\end{itemize}

The remainder of this paper is organized as follows: Section \ref{sec:litt} reviews related works. Section \ref{sec:dataset} details the process by which we created the dataset of code used in this study. In Section \ref{sec:code analysis}, we analyze this code using several tools to determine its security level. In Section \ref{sec:user}, we examine the intentions of the users who request code from the model. In Section \ref{sec:reprompt}, we perform an experiment in which the same prompts present in the Wildchat dataset are fed to different LLM models, in order to test the effectiveness of prompt engineering strategies on a real dataset.  Section \ref{sec:Refusal},  focuses on the cases in which the models refused to generate a program.  Section \ref{sec: Diss} discusses observations and insights that can be gleaned from our results. Concluding remarks are given in Section \ref{sec: Conclu}.

\section{Review of the Literature}
\label{sec:litt}
\vspace{-.5cm}
The rapid adoption of generative AI for software development is evidenced by the widespread use of tools such as GitHub Copilot~\cite{github2025survey} and Amazon CodeWhisperer~\cite{aws2024codewhisperer}. A recent GitHub survey reports that more than 90\% of developers now rely on generative AI to support their programming activities~\cite{github2025survey}. Despite these productivity gains, a growing body of empirical evidence shows that Large Language Models (LLMs) frequently generate insecure code or fail to adequately warn users about potential security risks~\cite{khoury2023secure,fu2025security,tihanyi2023new,spracklen2025we}. These findings raise serious concerns about the safety of deploying LLM-assisted programming tools in real-world development environments.

Two primary factors contribute to the prevalence of unsafe code generated by LLMs. First, widely used benchmarks for code generation rarely include constructs for evaluating security properties~\cite{bai2024apilot}. Second, existing evaluation metrics predominantly focus on functional correctness, performance, or usability, while largely ignoring security considerations~\cite{ashrafi2025enhancing,fakhoury2024llm,liu2023your,huang2024bias,yeticstiren2023evaluating}. Consequently, models that perform well on standard benchmarks may still produce code that is vulnerable in practice.

A growing line of research therefore focuses on explicitly evaluating the security of LLM-generated code. Early empirical studies demonstrated that functional correctness does not imply security: Khoury et al.~\cite{khoury2023secure} showed that ChatGPT-generated code frequently contains vulnerabilities when analyzed using static security checks. He and Vechev~\cite{he2023large} extended this work by integrating adversarial testing and security hardening techniques, revealing systematic weaknesses in LLM-generated code that can be exploited by attackers. Complementary large-scale evidence from production repositories further confirms these risks; Fu et al.~\cite{fu2025security} show that Copilot-generated code introduces security weaknesses into real GitHub projects.

Subsequent work has sought to formalize security-oriented evaluation frameworks for LLM-generated code. PromSec~\cite{nazzal2024promsec} and APILOT~\cite{bai2024apilot} incorporate static analysis and API-misuse detection into evaluation pipelines to capture security-relevant behaviors during generation. CoSec~\cite{li2024cosec} and ProSec~\cite{xu2024prosec} introduce learning- and decoding-time interventions that actively steer LLMs away from insecure patterns. More recently, CWEval~\cite{peng2025cweval} proposed an outcome-driven benchmark that jointly evaluates functionality and security using security-aware test oracles. In parallel, domain-specific studies have expanded security evaluation to package dependencies~\cite{spracklen2025we}, web applications~\cite{toth2024php,dora2025hidden}, and multi-language settings~\cite{kharma2025securityquality}.

Table~\ref{tab:llm_codegen_security} summarizes prior work that performs explicit security checking of LLM-generated code. As shown in the table, most studies rely primarily on static analysis tools, often in combination with limited dynamic testing or manual inspection to assess whether the code is secure. While these approaches have advanced the state of evaluation, they remain largely language-dependent and are typically applied to code generated by synthetic prompts or from controlled benchmarks ~\cite{he2023large,li2024cosec,bai2024apilot,nazzal2024promsec}.  Consequently, it is not clear that these studies are representative of real-world developer behavior.  Although some work uses Stack Overflow questions as a proxy for user intent~\cite{zeng2025inducing}, this approach risks pre-training bias, as LLMs may have encountered similar content during training. As a result, reported performance may be artificially inflated and fail to capture realistic failure modes.


Another challenge arises from representation and analysis techniques. Traditional NLP methods are insufficient for code, as program semantics differ substantially from natural language. To address this, prior studies commonly adopt structural program representations such as Abstract Syntax Trees, control-flow graphs, or data-flow graphs~\cite{bai2024apilot}. However, even with these representations, security evaluation remains tightly coupled to language-specific tooling~\cite{nazzal2024promsec,he2023large}, limiting generalizability across programming ecosystems.

In contrast to prior work that primarily evaluates the security of LLM-generated code in isolation, our study examines security awareness in authentic user interactions. Leveraging the WildChat dataset~\cite{zhao2024wildchat}, we conduct the first large-scale empirical investigation of how LLMs respond to insecure user-provided code in real conversations. By analyzing whether models proactively identify vulnerabilities and explain their causes, exploitation scenarios, and mitigations~\cite{sajadi2025llms}, we provide complementary insights into both model behavior and user intent. This perspective enables a more realistic understanding of the risks and real-world dynamics associated with LLM-assisted programming.

\begin{table*}[t]
\centering
\small
\renewcommand{\arraystretch}{1.15}
\resizebox{\textwidth}{!}{%
\begin{tabular}{p{5.2cm} p{3.6cm} p{3.8cm} p{6.4cm}}
\hline
\textbf{Paper} &
\textbf{Programming Language(s) Studied} &
\textbf{LLM Models Used} &
\textbf{Security Checking Tools / Methods} \\
\hline

Khoury et al.~(2023) -- \textit{How Secure Is Code Generated by ChatGPT?} &
Python, Java &
ChatGPT (GPT-3.5) &
Static vulnerability analysis, pattern inspection \\

He \& Vechev~(2023) -- \textit{Large Language Models for Code} &
Python, C/C++ &
GPT-3.5, Codex &
Static analysis, adversarial testing \\

Tihanyi et al.~(2023) -- \textit{Self-Healing Software via LLMs} &
C/C++, Java &
GPT-3.5 &
Formal verification, symbolic reasoning \\

Siddiq \& Santos~(2023) -- \textit{Generate and Pray} &
Python, Java &
GPT-3.5, GPT-4 &
CodeQL, Bandit, SpotBugs, CWE classification \\

Siddiq et al.~(2024) -- \textit{SALLM} &
Python, Java &
GPT-3.5, GPT-4 &
CodeQL, Bandit, SpotBugs, security evaluation framework \\

Bai et al.~(2024) -- \textit{APILOT} &
Python &
GPT-4 &
API misuse detection, static analysis \\

Tony et al.~(2024) -- \textit{Prompting Techniques for Secure Code Generation} &
Python, Java &
GPT-4 &
Static vulnerability scanning \\

Nazzal et al.~(2024) -- \textit{PromSec} &
Python, Java &
GPT-3.5, GPT-4 &
CodeQL, language-dependent SAST \\

Li et al.~(2024) -- \textit{CoSec} &
Python &
GPT-4 &
Supervised co-decoding, security rule enforcement \\

Xu et al.~(2024) -- \textit{ProSec} &
Python &
GPT-4 &
Proactive security alignment objectives \\

Nu{\~n}ez et al.~(2024) -- \textit{AutoSafeCoder} &
Python, JavaScript &
GPT-3.5, GPT-4 &
CodeQL, static analysis, fuzz testing \\

Kavian et al.~(2024) -- \textit{LLM Security Guard for Code} &
Python, Java &
GPT-3.5, GPT-4 &
CodeQL, Bandit, LLM-guided security repair \\

Bruni et al.~(2025) -- \textit{Benchmarking Prompt Engineering} &
Python &
GPT-4 &
Static security analysis, CWE mapping \\

Zeng et al.~(2025) -- \textit{Inducing Vulnerable Code Generation} &
Python, JavaScript &
GPT-4 &
Manual vulnerability inspection \\

Fu et al.~(2025) -- \textit{Security Weaknesses of Copilot-Generated Code} &
Python, Java &
GitHub Copilot &
Static security analysis, vulnerability mining \\

Spracklen et al.~(2025) -- \textit{Package Hallucinations} &
Python &
GPT-4 &
Dependency hallucination and security analysis \\

Peng et al.~(2025) -- \textit{CWEval} &
Python, Java, JavaScript &
GPT-3.5, GPT-4 &
CWEval benchmark, security oracles \\

Kharma et al.~(2025) -- \textit{Security \& Quality in LLM-Generated Code} &
Python, Java, JavaScript, C++ &
GPT-3.5, GPT-4, Claude, LLaMA &
CodeQL, Bandit, Semgrep \\

Dora et al.~(2025) -- \textit{Hidden Risks of LLM-Generated Web Code} &
JavaScript, Python, PHP &
GPT-4, Claude, Gemini &
OWASP Top-10 checklist, expert review \\

T{\'o}th et al.~(2024) -- \textit{LLMs in Web Development (PHP)} &
PHP &
GPT-4 &
Burp Suite (DAST), PHP static analyzers \\

Sajadi et al.~(2025) -- \textit{Do LLMs Consider Security?} &
Python, JavaScript &
GPT-4, Claude~3, LLaMA~3 &
CodeQL (ground truth), manual analysis of security warnings \\

\hline
\end{tabular}}
\caption{Summary of prior studies that perform explicit security checking of LLM-generated code or LLM responses to code.}
\label{tab:llm_codegen_security}
\end{table*}

\section{Construction of the Dataset}
\label{sec:dataset}

\begin{figure}[]
\centering
\includegraphics[width=1\columnwidth]{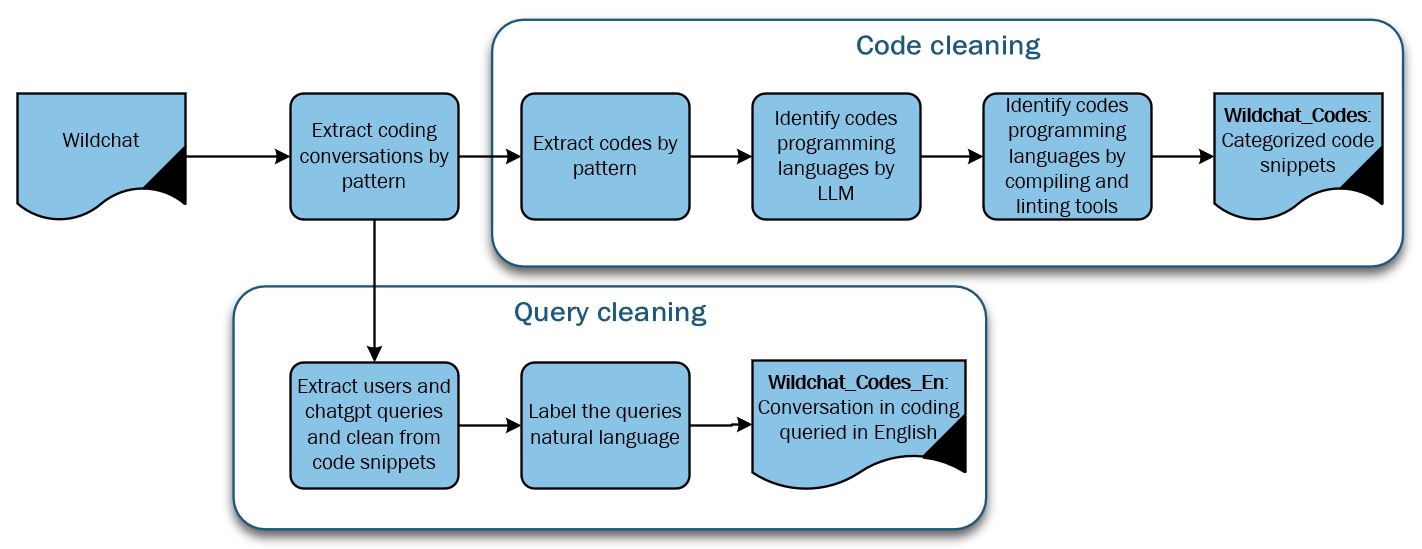}
\caption{Dataset generation pipeline}
\label{fig:dataset_pipeline}
\end{figure}

The basis for this study is the conversations related to the code or coding tasks present in the WildChat data set \cite{zhao2024wildchat}. WildChat is a database of 1 million real-life conversations with different versions of ChatGPT collected between April 2023 and May 2024.  We extracted the relevant codes and conversations from this dataset using the process illustrated in Figure \ref{fig:dataset_pipeline}. The WildChat database is ideal for academic research because of it’s scale, and because it is anonymized, multilingual, and labelled. 

Each conversation in WildChat is identified by a unique \textit{conversation\_hash} and consists of a sequence of user queries and ChatGPT responses within a session. As an initial step, we extracted the subset of conversations that contain code. In the dataset, code snippets are delimited by a single backtick (\textasciigrave) or by triple backticks (\textasciigrave\textasciigrave\textasciigrave), which are often, but not always, followed by the programming language used. Snippets enclosed in a single backtick are typically short fragments, often a single line of code, whereas those enclosed in triple backticks are longer and more complex code snippets. For the purpose of our analysis, we concentrated on the latter. Of a total of 837,989 conversations in the WildChat dataset, 82,843 contain code generated by ChatGPT. These code-containing conversations span multiple model versions, including \texttt{gpt-4-0314}(6.4\%), \texttt{gpt-3.5-turbo-0301}(23.3\%), \texttt{gpt-3.5-turbo-0613}(44.3\%), \texttt{gpt-4-1106-preview}(12\%), \texttt{gpt-3.5-turbo-0125}(6.9\%), and \texttt{gpt-4-0125-pr eview} (7.4\%). We conducted the analysis presented in this paper separately for each of these model versions, and the outcomes were found to be largely consistent across all of them. Therefore, to avoid redundancy and repetition of similar concepts and results, we report the findings in a cumulative manner that represents the aggregated behaviour across models.

In the WildChat conversations, code snippets delimited by triple backticks usually begin with a programming language tag, followed by the code itself. However, these tags are often missing or incorrectly specified, so they cannot be relied on. To handle snippets without valid annotations, we applied a programming language identification model.\footnote{\url{https://huggingface.co/philomath-1209/programming-language-identification}} to automatically classify the code snippets by language. The model supports 26 programming languages with a reported accuracy of 95\%, which implies that up to 5\% of labels may still be incorrect. To further enhance labelling quality, we validated snippets for the six programming languages (those above the middle line in Table~\ref{tab:codestat}) using language-specific syntax checkers: we employed the py\_compile module\footnote{\url{https://docs.python.org/3/library/py\_compile.html}}
 for Python, eslint\footnote{\url{https://github.com/eslint}}
 for JavaScript, javac for Java, gcc for C and C++, the php -l command for PHP, and Microsoft's Roslyn compiler platform\footnote{\url{https://learn.microsoft.com/en-us/dotnet/csharp/roslyn-sdk/get-started/syntax-analysis}}
 for C\#. This validation ensured syntactic correctness and improved the accuracy of the labelling, which is essential for reliable downstream analysis. The resulting coding fragments, labelled by programming language,  constitute the dataset named \textit{WildChat\_Codes}, as illustrated in Figure~\ref{fig:dataset_pipeline}.


As noted above, we also investigated users’ coding intentions and their inquiries about coding issues. Since code-related conversations were not always conducted in English, it was necessary to first isolate English conversations for analysis. The WildChat dataset provides a `Language' column; out of 82,843 conversations that contain code,  48,391 are labelled as being in English. However, we observed that many of these were, in fact, not in English, particularly when the conversation text contained numerous short code snippets (see, for example, the conversation with \texttt{conversation\_hash  e8e1274f7c1253299fa6c7865ee45703}). To ensure reliable language identification, we removed all code snippets from the conversation text and subsequently applied the Python langdetect library to uncover the true language of each conversation. Among the 82,843 conversations that contain code, 34,478 were identified as having queries in English. We focused on English conversations due to the broader availability of language processing tools and to enable the authors to manually verify and double-check the results with confidence. These English coding conversations constitute the dataset named \textit{WildCode\_EN}, as illustrated in Figure~\ref{fig:dataset_pipeline}.

Each conversation consists of a sequence of pairs: each pair consists of a user query followed by a ChatGPT reply. To study the intentions of users and their interactions with ChatGPT’s responses, we refer to the first user query as the \textbf{initial query}, with all subsequent user queries referred to as \textbf{follow-up queries}. Similarly, the first ChatGPT response is defined as the \textbf{initial response}, and all subsequent replies are considered \textbf{follow-up responses}.

\section{Code Analysis}
\label{sec:code analysis}
\subsection{Overview of the Code Snippets}
\begin{table}[ht]
\caption{Code stats}

\begin{tabularx}{\textwidth}{XXXXXXX}
\toprule
\textbf{Language} & \textbf{Code Snippets} & \textbf{Conv.} & \textbf{Avg Snippets/Conv.} & \textbf{Avg. Lines} & \textbf{Stddev. Lines} & \textbf{Avg Comments} \\
\midrule
C/CPP            &7,526&     3,911&      1.92&     43.73&     38.61&       3.18\\
C\#           &14,138&     5,895&      2.40&     28.39&     26.81&      2.27\\
Java         &18,680&     7,228&      2.58&     27.43&     27.89&       1.94\\
JavaScript   &15,943&     7,217&      2.21&     23.39&     25.15&       2.23\\
PHP           &449&     340&      1.32&     20.60&     13.42&       3.05\\
Python       &60,451&    22,949&      2.63&     26.26&     24.75&       3.44\\
\midrule
Rust          &1,919&     1,001&      1.92&     19.15&     17.46&       1.78\\
COBOL         &1,378&     1,022&      1.35&     15.84&     20.00&       0.06\\
Fortran         &679&       477&      1.42&     15.40&     17.83&       0.42\\
jq            &2,805&     1,146&      2.45&     14.23&      6.46&       0.06\\
Ruby          &1,507&     1,097&      1.37&     14.30&     16.34&       1.31\\
AppleScript      &192&       128&      1.50&     16.36&     17.89&       1.45\\
Kotlin        &1,559&       580&      2.69&     22.95&     20.45&       1.79\\
ARM Assembly    &1,174&       804&      1.46&     22.91&     28.68&       2.74\\
Erlang        &2,225&     1,491&      1.49&     15.36&     17.55&       0.59\\
Swift           &819&       552&      1.48&     14.77&     12.82&       1.11\\
R             &1,555&       730&      2.13&     13.94&     13.02&       2.38\\
PowerShell    &6,375&     2,856&      2.23&     16.68&     10.47&       0.49\\
Scala         &1,692&     1,111&      1.52&     13.33&     11.87&       0.94\\
Lua             &401&       288&      1.39&     12.34&      9.95&       2.05\\
Pascal        &3,623&     2,325&      1.56&     17.01&     17.37&       0.72\\
Go            &1,631&       768&      2.12&     35.49&     34.43&       2.89\\
Perl          &1,692&     1,255&      1.35&     12.55&     13.05&       0.84\\
Wolfram       &1,225&       783&      1.56&     15.80&     15.86&       1.20\\
.NET          &3,707&     1,637&      2.26&     27.45&     20.47&       4.49\\

\bottomrule
\end{tabularx}
\label{tab:codestat}
\end{table}

Table~\ref{tab:codestat} presents detailed statistics on code snippets extracted from the WildCode (shown in Figure \ref{fig:dataset_pipeline}) dataset, broken down by programming language. In this context, each code snippet within a conversation has been analyzed independently, and a single conversation may contain multiple code snippets. The table reports the average number of snippets per conversation, the average number of lines per snippet, the average number of comments per snippet, and the standard deviation of the number of lines per snippet.

As can be seen in  Table \ref{tab:codestat}, Python dominates with 60,451 code snippets across 22,949 conversations, making it by far the most common language in the dataset. C/C++ code snippets are the longest, averaging 43.7 lines per code snippet with high variability (std. dev. 38.6). Moreover, among all languages, .NET has the highest average comments per block (4.49), and Python also ranks high (3.44), reflecting their widespread use and strong interest among users. The results show that LLMs predominantly generate short programs, though with substantial variation in length. Popular programming languages exhibit both a higher frequency of code blocks and a greater density of comments within the code. On average, each conversation contains between 1.5 and 2.5 code snippets, indicating that users frequently prompt the LLM for iterative refinements of previously generated code. The Intent section of this paper examines this phenomenon in greater detail and analyzes the corresponding follow-up categories.

\subsection{Syntax Check}
As explained in Section \ref{sec:dataset}, we conducted a syntax check on a subset of code snippets in WildCode. This subset includes code written in the six programming languages, as listed above the dividing line in Table~\ref{tab:codestat}. More specifically, only the snippets that were not labelled by ChatGPT and had to be labelled by a classifier.

Table~\ref{tab:syntax_summary} reports the number of code snippets labelled by the model for each language. These correspond to the snippets for which ChatGPT did not initially provide a valid language tag in the \texttt{```LANGUAGE} format, and thus required the classification model to determine their language. For each language, we also report the subset of snippets containing syntax errors identified by the linting process, and from this, we infer the number of valid snippets, i.e., those without syntax errors. 
Note that no single PHP code snippet was valid, as none started with the \texttt{<?php} tag. A manual inspection further showed that the classified code snippets were all related to command-line tools associated with PHP rather than actual PHP source code.

\begin{table}[]
\centering
\caption{Number of snippets labelled by the model and valid (syntax-error-free) messages per language.}
\label{tab:syntax_summary}
\small
\setlength{\tabcolsep}{6pt}
\renewcommand{\arraystretch}{1.2}
\begin{tabularx}{\textwidth}{XXXXX}
\toprule
\textbf{Language} & \textbf{Labelled by ChatGPT} & \textbf{Labelled by Model} & \textbf{Codes w. Syntax Err.} & \textbf{Valid Code Snippets} \\
\midrule
Python     & 60,451 & 57,371 & 19,805 & 37,566 \\
Java       & 18,680 & 19,257 & 20     & 19,237 \\
JavaScript & 15,943 & 26,442 & 15,725 & 10,717 \\
C\#        & 14,138 & 13,893 & 6,517  & 7,376 \\
C/CPP      & 7,526 & 21,510 & 18,323 & 3,187 \\
PHP        & 449 & 1,935  & 1,935  & 0 \\
\bottomrule
\end{tabularx}
\end{table}

To classify linting messages into syntax error categories, we first embedded the natural-language descriptions of the 20 predefined categories\footnote{\url{https://github.com/regularpooria/WildCode/blob/master/utils/error\_categories.json}}. Each linting message was then embedded and assigned to the category with the highest cosine similarity to its description embedding. Table~\ref{tab:syntax_categories_5clusters} presents the distribution of messages in categories and programming languages. We observe that C/C++ and C\# dominate categories related to general programming issues (e.g. parsing errors or missing semicolons), while JavaScript, Python, and PHP contribute primarily to language-specific parsing errors.

\begin{table}[ht]
\centering
\caption{Aggregated syntax error categories across languages (total $177{,}732$ rows).}
\label{tab:syntax_categories_5clusters}
\small
\setlength{\tabcolsep}{6pt}
\renewcommand{\arraystretch}{1.2}
\begin{tabularx}{\textwidth}{l X X X X X X X}
\toprule
\textbf{Error Category} & \textbf{Mess.} & \textbf{C/C++} & \textbf{C\#} & \textbf{Java} & \textbf{JS} & \textbf{Python} & \textbf{PHP} \\
\midrule
Syntax Error       & 95,432 & 48,598 & 30,032 & 0 & 7,212 & 9,585 & 0 \\
Declaration Error       & 33,026 & 15,683 & 11,013 & 0 & 2,801 & 3,529 & 0 \\
Access Control Error       & 20,852 & 8,106 & 7,837 & 0 & 2,168 & 2,741 & 0 \\
Preprocessor Error       & 18,377 & 5,390 & 5,172 & 100 & 2,759 & 3,021 & 1935 \\
Lexical Error          & 2,829 & 1,244 & 885 & 0 & 323 & 377 & 0 \\
\bottomrule
\end{tabularx}
\end{table}
\subsection{Security Analysis}
We used OpenGrep\footnote{\url{https://github.com/opengrep/opengrep}}, an open source tool to analyze the security issues of the code in our WildCode dataset. There is a repository of rules for OpenGrep\footnote{\url{https://github.com/regularpooria/opengrep-rules}} that includes security-related rules. These rules are defined using regular expressions that identify common insecure coding patterns across multiple programming languages. For our study, we selected the subset of security rules for the six programming languages represented in the WildCode dataset, resulting in a total of 648 rules. Each rule has a CWE mapping. The distribution of these rules across programming languages is presented in Table~\ref{tab:vuln_categories} in column \texttt{Rules} for each category and for each language, and the list of rules is available in this file\footnote{\url{https://github.com/regularpooria/WildCode/blob/master/utils/rules.json}} on our GitHub.

\begin{table}[ht]
\centering
\caption{Possible vulnerabilities by category and language.}
\label{tab:vuln_categories}
\small
\setlength{\tabcolsep}{4pt}
\renewcommand{\arraystretch}{1.2}
\begin{tabularx}{\textwidth}{
    l
    *{12}{>{\centering\arraybackslash}X}
}
\toprule
\multirow{2}{*}{Language} &
\multicolumn{3}{c}{Hash Function} &
\multicolumn{3}{c}{SQL Injection} &
\multicolumn{3}{c}{RNG} &
\multicolumn{3}{c}{Deserialization} \\
\cmidrule(lr){2-4} \cmidrule(lr){5-7} \cmidrule(lr){8-10} \cmidrule(lr){11-13}
& CS & TO & Rules
& CS & TO & Rules
& CS & TO & Rules
& CS & TO & Rules \\
\midrule
C\#        & 24  & 56   & 3   & 124  & 624   & 1   & 91   & 133   & 1   & 56   & 82    & 11  \\
Java       & 32  & 93   & 16  & 223  & 1373  & 11  & 150  & 232   & 1   & 273  & 634   & 15  \\
JavaScript & 22  & 72   & 4   & 163  & 932   & 9   & 352  & 987   & 1   & 463  & 805   & 5   \\
PHP        & 17  & 73   & 5   & 183  & 873   & 4   & N/A  & N/A   & N/A & N/A  & N/A   & N/A \\
Python     & 180 & 577  & 22  & 883  & 6985  & 17  & 2810 & 12791 & 3   & 2258 & 4833  & 8   \\
\bottomrule
\end{tabularx}
\begin{tablenotes}
\small
\item CS: Code Snippets | TO: Total occurrences | Rules: Number of detection rules
\end{tablenotes}
\end{table}

Table~\ref{tab:vuln_categories} presents the distribution of code snippets mapped to regular expression rules in four vulnerability categories: hash functions, SQL injection, random number generation (RNG) and deserialization. In the table~\ref{tab:vuln_categories}, \textbf{TO} (Total Occurrences) denotes the total number of rule matches, while \textbf{CS} (Code Snippets) indicates the number of unique code snippets in which at least one rule from the corresponding category was identified. Because a single code snippet can match (i.e., violate) multiple rules, the values in \textbf{TO} are always greater than or equal to those in \textbf{CS}. Note that entries marked N/A indicate that no corresponding rules exist for that programming language and that C/C++ does not apply to the table. These rules are detailed in the following subsections. 

\textbf{Note:} In tables \ref{tab:hash_vulnerabilities}, \ref{tab:sql_vulnerabilities}, \ref{tab:unsafe_deserialization}, \ref{tab:unsafe_memory}; The sum of unique conversation hashes across individual rules may be greater than the overall unique hashes for the language, because some conversation hashes violate multiple rules.



\subsubsection{Weak Cryptographic hash functions:}
A total of 264 unique ChatGPT-generated conversations contain code snippets referencing hash functions across Python, Java, C, C\#, JavaScript, and PHP. Using a set of 50 static analysis rules tailored to detect insecure or improper use of hash functions, we found that 54 conversations triggered at least one rule, representing a vulnerability rate of 20.61\%. Notably, only 13 out of the 50 rules were activated, with the majority of violations stemming from the continued use of MD5, SHA1, or cryptographic algorithms lacking authentication guaranties. Table \ref{tab:hash_vulnerabilities} shows all vulnerabilities found in the conversations.

\begin{table}[ht]
\centering
\begin{threeparttable}
\caption{Hash Function Vulnerabilities in ChatGPT-Generated Codes By Language}
\label{tab:hash_vulnerabilities}
\begin{tabularx}{\textwidth}{l X X r r}
\toprule
\textbf{Language} & \textbf{Rule} & \textbf{CWE} & \textbf{Unique Files} & \textbf{Total Occurrences} \\
\midrule
\multirow{7}{*}{Java} 
  & use-of-md5               & CWE-328 & 7  & 8  \\
  & desede-is-deprecated     & CWE-326 & 1  & 1  \\
  & des-is-deprecated        & CWE-326 & 1  & 2  \\
  & ecb-cipher               & CWE-327 & 1  & 2  \\
  & use-of-aes-ecb           & CWE-327 & 1  & 2  \\
  & use-of-default-aes       & CWE-327 & 1  & 2  \\
  & \textbf{Overall}         &  & \textbf{10 (31.25\%)} & \textbf{17} \\
\midrule
\multirow{3}{*}{PHP} 
  & weak-crypto              & CWE-328 & 2  & 2  \\
  & openssl-decrypt-validate & CWE-252 & 1  & 1  \\
  & \textbf{Overall}         &  & \textbf{3 (17.65\%)}  & \textbf{3} \\
\midrule
\multirow{6}{*}{Python} 
  & insecure-hash-algorithm-md5         & CWE-327 & 33 & 59 \\
  & insecure-hash-algorithm-sha1        & CWE-327 & 4  & 6  \\
  & crypto-mode-without-authentication  & CWE-327 & 4 & 5  \\
  & insecure-cipher-algorithm-des       & CWE-327 & 1  & 2  \\
  & md5-used-as-password                & CWE-327 & 2  & 2  \\
  & \textbf{Overall}                    &  & \textbf{41 (22.78\%)} & \textbf{74} \\
\bottomrule
\end{tabularx}
\end{threeparttable}
\end{table}

\subsubsection{SQL Injection:}
We examined 970 LLM-generated conversations that contained SQL-related code snippets through regex patterns, applying 42 rules targeting common patterns of SQL injection vulnerabilities. Only seven rules were triggered, resulting in 61 logged conversations, a vulnerability rate of 3.93\%. The most frequently violated rule involved the execution of raw SQL queries in SQLAlchemy, followed by tainted string concatenation and JDBC usage patterns. Table \ref{tab:sql_vulnerabilities} shows the rules and their occurrences.

\begin{table}[!ht]
\centering
\begin{threeparttable}
\caption{SQL Injection Vulnerabilities in ChatGPT-Generated Codes By Language}
\label{tab:sql_vulnerabilities}
\begin{tabularx}{\textwidth}{l X X r r}
\toprule
\textbf{Language} & \textbf{Rule} & \textbf{CWE} & \textbf{Code Snippets} & \textbf{Total Occurrences} \\
\midrule
\multirow{2}{*}{C\#} 
  & csharp-sqli      & CWE-89 & 3  & 3  \\
  & \textbf{Overall} &        & \textbf{3 (2.42\%)} & \textbf{3} \\
\midrule
\multirow{2}{*}{Java} 
  & jdbc-sqli       & CWE-89 & 4  & 9  \\
  & \textbf{Overall} &  & \textbf{4 (1.79\%)} & \textbf{9} \\
\midrule
\multirow{2}{*}{JavaScript} 
  & tainted-sql-string & CWE-915     & 2  & 3  \\
  & \textbf{Overall}   &      & \textbf{2 (1.23\%)} & \textbf{3} \\
\midrule
\multirow{2}{*}{PHP} 
  & tainted-sql-string & CWE-915     & 7  & 9  \\
  & \textbf{Overall}    &     & \textbf{7 (12.50\%)} & \textbf{9} \\
\midrule
\multirow{6}{*}{Python} 
  & tainted-sql-string   &  CWE-915  & 2 & 7 \\
  & sqlalchemy-execute-raw-query   &CWE-89   & 42  & 70  \\
  & psycopg-sqli & CWE-89 & 5 & 5  \\
  & sql-injection-db-cursor-execute &  CWE-89   & 3  & 15  \\
  & avoid-sqlalchemy-text &  CWE-89            & 2  & 2  \\
  & \textbf{Overall} &                 & \textbf{45 (5.40\%)} & \textbf{99} \\
\bottomrule
\end{tabularx}
\end{threeparttable}
\end{table}

\subsubsection{Weak random number generation:}
We examined 3032 conversations that contained code that uses random number generation. If the generated random number is used for a security-sensitive task, such as creating a password or a cryptographic nonce, then the underlying random number generation algorithm must be \texttt{cryptographically secure}. Otherwise, a vulnerability is present in the code. We applied this analysis to Java, C\#, JavaScript, PHP and Python, and followed by a manual review of the detected samples to confirm their context and correctness. We found 17 instances where weak random number generation was found.  Of these,  15 were present in Java  code and 2 in Python. Thus, only 0.47\% of the code snippets contained this specific vulnerability, a result that is significantly better than the one obtained for other classes of vulnerabilities.

\subsubsection{Deserialization attacks:}
A deserialization vulnerability is present whenever the code processes serialized data from an untrusted source, without including proper validation and security checks. This gives a malicious adversary the ability to input arbitrary data, perform a denial of service, or even execute arbitrary code \cite{sayar2023depth}. This is one of the main security issues in Java programs, a fact that was made evident by the devastating Log4Shell vulnerability in 2021 \cite{log4shell}. 

Java programs present in the Wildchat dataset include 30 instances of deserialization. A manual inspection revealed that every single case seemed vulnerable to deserialization attacks, as none contained security checks. Furthermore, in none of the conversations associated with these programs did ChatGPT discuss the risks inherent in deserializing data. 
\begin{table}[th]
\centering
\begin{threeparttable}
\caption{Unsafe Deserialization in ChatGPT-Generated Codes By Language}
\label{tab:unsafe_deserialization}
\begin{tabularx}{\textwidth}{l X l r r}
\toprule
\textbf{Language} & \textbf{Rule} & \textbf{CWE} & \textbf{Scripts} & \textbf{Total Occurrences} \\
\midrule
\multirow{6}{*}{Java} 
  & documentbuilderfactory-disallow-doctype-decl-missing      & CWE-611 & 3  & 3  \\
  & object-deserialization      & CWE-502 & 21  & 33  \\
  & transformerfactory-dtds-not-disabled      & CWE-611 & 1  & 1  \\
  & saxparserfactory-disallow-doctype-decl-missing      & CWE-611 & 5  & 7  \\
  & use-snakeyaml-constructor      & CWE-502 & 1  & 3  \\
  & \textbf{Overall} &        & \textbf{30 (10.99\%)} & \textbf{47} \\
\midrule
\multirow{2}{*}{JavaScript} 
  & grpc-nodejs-insecure-connection       & CWE-502 & 7  & 20  \\
  & \textbf{Overall} &  & \textbf{7(1.51\%)} & \textbf{20} \\
\midrule
\multirow{2}{*}{Python} 
  & marshal-usage       & CWE-502 & 2  & 3  \\
  & \textbf{Overall} &  & \textbf{2(0.09\%)} & \textbf{3} \\
\bottomrule
\end{tabularx}

\end{threeparttable}
\end{table}


\vspace{-10pt}
\subsubsection{Memory Safety:}

In our previous paper \cite{khoury2023secure}, ChatGPT exhibited particular difficulty with memory corruption vulnerabilities in C/C++ programs. This is also the case for programs present in the Wildchat data set.

\begin{table}[ht]
\centering
\begin{threeparttable}
\caption{Unsafe Memory in ChatGPT-Generated Codes By Language}
\label{tab:unsafe_memory}
\begin{tabularx}{\textwidth}{l X X r r}
\toprule
\textbf{Language} & \textbf{Rule} & \textbf{CWE} & \textbf{Code Snippets} & \textbf{Total Occurrences} \\
\midrule
\multirow{7}{*}{C/CPP} 
  & insecure-use-scanf-fn      & CWE-676 & 378  & 1182  \\
  & insecure-use-memset      & CWE-14 & 117  & 263  \\
  & insecure-use-string-copy-fn      & CWE-676 & 120  & 273  \\
  & insecure-use-strcat-fn      & CWE-676 & 18  & 56  \\
  & insecure-use-gets-fn      & CWE-676 & 11  & 19  \\
  & insecure-use-printf-fn      & CWE-134 & 5  & 14  \\
  & \textbf{Overall} &        & \textbf{581 (14.85\%)} & \textbf{1807} \\
\bottomrule
\end{tabularx}

\end{threeparttable}
\end{table}

We focus on 6 rules, which forbid the use of constructs that are known to be easily exploitable, namely \texttt{scanf},   \texttt{strcpy},   \texttt{memset},   \texttt{strcat},   \texttt \texttt{gets} \texttt{printf}.   Using any of these functions incurs a risk of a buffer overflow, unless accompanied by thorough boundary checks, or unless the input is completely controlled by the program.    As shown in Table \ref{tab:unsafe_memory}, the C/C++ programs combine 1807 distinct violations of these 6 rules, across 581 distinct code snippets. In general, 14.85\% of the C/C++ programs contain at least 1 violation. Many code fragments contained multiple occurrences, often several dozen distinct occurrences.   The continued use of these function calls is particularly disappointing, since in most cases memory-safe alternatives exist.

This is almost certainly an undercount of the actual number of vulnerable programs, since even in the absence of these functions, memory corruption can still occur because of errors in pointer arithmetic or memory management.  We also found that programs that do not contain memory management errors are substantially shorter than those that do, with an average of 48 lines in the former case (Std. dev. 40) versus 106 (std. dev. 58) in the latter case. That said, a program of 100 lines is hardly a ``large'' program by any definition, and the inability of the model to create a memory-safe program of even such a short size is disheartening. 

Interestingly, despite the fact that multiple C-rules are triggered hundreds of times, 3 rules are never triggered: a rule forbidding freeing a pointer twice and two rules related to use-after-free of pointers.  Violations of these rules usually occur in larger code-bases, when the same pointer variable is used in several different functions or in several different files. LLMs are still limited in the size of the programs they can produce, and these two types of vulnerabilities can usually be easily avoided when writing a small code fragment. If a novice programmer intends to create a larger program by separately requesting several fragments from the LLMs and joining them together, it is likely that these vulnerabilities may be present in the final code.

\subsection{ReDoS (Regular Expression Denial of Service)}

In our previous research \cite{khoury2023secure}, one of the security issues for which ChatGPT  seemed to struggle the most is the problem of ReDoS attacks \cite{owasp-redos}. It is one of a handful of cases in which the model is unable to recognize the presence of the vulnerability even when explicitly prompted on the topic.

To estimate the prevalence of vulnerable regular expressions in code generated by ChatGPT, we extracted every regex from the programs in our dataset. We then analyzed these expressions using four different reDoS vulnerability detection tools. The tools used are:\texttt{Saferegex} \cite{SafeRegex}, \texttt{Rescue} \cite{Recue}, \texttt{Redoshunter} \cite{redoshunter} and \texttt{Revealer} \cite{revealer}. 


According to this analysis, about a third of the regex present in the code in our dataset is susceptible to ReDoS attacks. Despite the possibility that these tools may generate false positives, this analysis is likely an overcount of the actual incidence of ReDoS vulnerabilities in the generated code snippet. Code can safely use a vulnerable regex if it does not manipulate untrusted user input, if the regex validation algorithm is not susceptible to ReDoS attacks, or if the validation is bounded by a timeout. However, this analysis provides us with a baseline indication of the prevalence of ReDoS vulnerabilities in our dataset.

The results are presented in Table~\ref{tab:regex}. For each programming language, we report the number of extracted regexes and the vulnerabilities identified by each detection tool. Note that Rescue can either mark a regex as vulnerable or Time Out, a result that indicates a likely but not certain vulnerability, while Revealer marks a vulnerable regex as either Polynomial or exponential. 

Only two English-language conversations explicitly mentioned the ReDoS attack, and never in the context of writing secure code.  In one case, a user reproduced a \texttt{npm} report which referred to CVE-2022-3517 (a ReDoS vulnerability) and asked how to respond to it. In the second case, ReDoS was part of a list of 100 vulnerability types for bug bounty programs. There was no instance of an LLM commenting that a regex (either produced by the user or by itself) is susceptible to this class of attack.


\begin{table}[ht]
\caption{Detection of Vulnerable Regexes by Different Tools.}
\label{tab:regex}
\small
\setlength{\tabcolsep}{4pt}
\renewcommand{\arraystretch}{1.2}
\begin{tabularx}{\textwidth}{
    X r r r r r r r r
}
\toprule
Lang. &  Regex   & SafeRegex &
\multicolumn{2}{c}{Rescue} &
ReDoSHunter &
\multicolumn{2}{c}{Revealer} &
Total Vuln. \\
\cmidrule(lr){3-3}\cmidrule(lr){4-5}\cmidrule(lr){6-6}\cmidrule(lr){7-8}
& Count& & Vuln. & T.O. & & Poly. & Expo. & \\
\midrule
C/CPP     & 80   & 40   & 2  & 14  & 10  & 1 & 0 & 36 (45\%) \\
C\#       & 46   & 27   & 0  & 5   & 5   & 0 & 0 & 27 (58.7\%) \\
Java      & 226  & 160  & 2  & 37  & 19  & 2 & 0 & 47 (20.7\%) \\
JS        & 50   & 16   & 2  & 9   & 11  & 1 & 2 & 16 (32\%) \\
PHP       & 38   & 11   & 0  & 8   & 1   & 0 & 0 & 11 (28.9\%) \\
Python    & 753  & 309  & 20 & 290 & 222 & 10 & 0 & 212 (28.1\%) \\
\midrule
Total     & 1,203 & 568 & 26 & 375 & 268 & 14 & 2 & 354 (29.4\%) \\
\bottomrule
\end{tabularx}
\end{table}

\subsection{Hallucinations}
\label{sec:hallucinations}


An important security vulnerability in code generated by LLMs is the persistent presence of 'package hallucinations', package names that the LLM includes in the code, but do not actually exist. This opens pathways for malicious adversaries to exploit the program by creating libraries with the names hallucinated by the LLM, and including malicious code in those libraries. Alternatively, the code may require manual modifications in order to run properly, with the attendant risk of introducing vulnerabilities, as discussed in \cite{khoury2023secure,spracklen2025we}. 

\subsubsection{Methodology}
\begin{figure}[ht]
\centering
\includegraphics[width=\columnwidth]{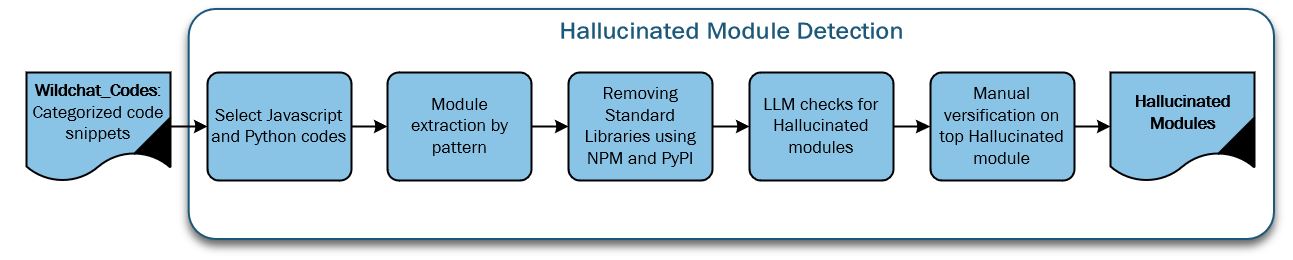}
\caption{Hallucination detection pipeline}
\label{fig:hallucination_pipeline}
\end{figure}

We focus our analysis on Python and JavaScript due to their widespread adoption and the strong ecosystem support provided by package repositories, PyPI for Python and NPM for JavaScript. These repositories allow for a systematic validation of the existence of third-party modules. Our pipeline for detecting hallucinated modules, illustrated in Figure \ref{fig:hallucination_pipeline}, begins by extracting imported module names using a regular expression. We then filter these candidates against the official PyPI and NPM package lists, discarding any modules not present in the repositories. Standard library modules are similarly excluded. Finally, as an additional safeguard, we query an LLM with Web Search capabilities to flag potentially invalid modules, which we manually review and remove if confirmed to be incorrect. We run this analysis on conversations found in Wildchat, which mostly used the \texttt{GPT3.5} model. We further fed the Wildchat prompts that led to the creation of  Python and JavaScript  to the newer models found in Table \ref{tab:reprompting_llms}, and examine the code these models returned as part of this analysis.  These models include open source OpenAI and Cohere models, and the analysis thus illustrates the difference between older and newer LLMs with respect to library hallucinations

\subsubsection{Results on WildChat (Older models)}
Our analysis of the conversations present in the Wildchat dataset identified 285 distinct hallucinated Python modules among 1,984 modules used in ChatGPT responses (approximately 14.4\%), with 210 (73.7\%) appearing only once. In comparison, 21 distinct JavaScript packages were hallucinated out of 606 (approximately 3.5\%), with 13 appearing only once. This pattern indicates that the model predominantly generates unique fictitious modules in Python, particularly when handling less familiar coding tasks. Interestingly, module names that occur in hundreds of code snippets are never hallucinations.  Instead, the model tends to produce imaginary modules when faced with less common or user-defined functionalities. Each hallucinated module name occurs fewer than 50 times in the dataset. 

\subsubsection{Results with newer models}
For this analysis, we used the prompts present in the Wildchat dataset that resulted in the creation of vulnerable code. Although this subset does not represent the Wildchat dataset in full, it provides meaningful insights as to  how newer LLMs can recall library names.

Our analysis identified 47 distinct hallucinated Python modules among 403 modules used across the responses of all 5 models, with 132 appearing only once. In comparison, 21 distinct JavaScript packages were hallucinated out of 325, with all 21 packages appearing only once. This pattern supports the idea that  newer models exhibit a lower amount of hallucinations and are making mistakes on one-off packages rather than collectively hallucinating a singular package.



\subsection{Secrets}
\label{sec:secrets}

\begin{figure}[H]
\centering
\includegraphics[width=\linewidth]{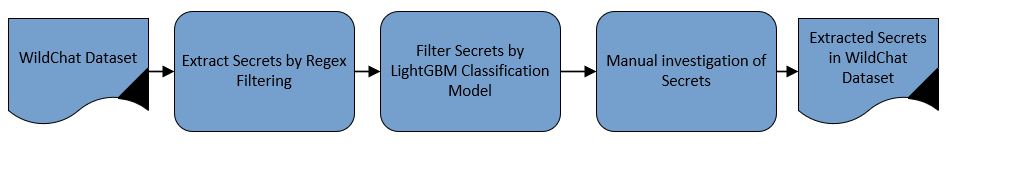}
\caption{Pipeline for the secret detection}
\label{fig:secret_pipeline}
\end{figure}

We analyze whether user prompts in WildCode contain exposed secrets, such as API keys or other credentials. Figure \ref{fig:secret_pipeline} shows the overall pipeline for detecting and removing placeholder secrets from conversations. The pipeline uses a combination of regex and a LightGBM model to perform this sanitization of the data.

\subsubsection{Regex Filtering}

In order to detect such secrets, we first filter the WildCode conversations with regex to find any parts that resemble an API key. A full list of regex strings that we use is available in our GitHub, but they include strings that start with "CLIENT\_SECRET", "SECRET", or similar strings. After the first pass, we get 1{,}380 conversations that contain a secret. Through manual inspection, we realized that many of the strings were placeholders inserted by the AI model, for the user to fill-in later. In order to differentiate between these strings, we decided to train a simple model to detect placeholders vs. real secrets through features found in the text.

\subsubsection{LightGBM Model}
In order to train the  model, we created a balanced synthetic dataset of placeholder and real secrets generated through a set of hard-coded rules and patterns that are randomly put together to resemble a secret.

\paragraph{Synthetic dataset construction.}
The dataset consists of 100{,}000 samples composed of:
\begin{itemize}
    \item 50{,}000 realistic, high-entropy tokens that mimic real-world secret formats.
    \item 50{,}000 placeholder or obviously fake tokens.
\end{itemize}

Real keys include patterns resembling AWS-style keys, GitHub tokens, Stripe keys, JWT-like structures, UUIDs, hex digests, base64-encoded strings, and randomly generated high-entropy sequences. Placeholder keys include literal strings such as \texttt{password}, \texttt{API\_KEY}, \texttt{insert\_your\_key}, templating expressions, repetitive patterns, and weak instructional examples.
To introduce variability and prevent over-reliance on specific features, each sample has a 10\% probability of random character mutation.

\begin{figure}[h]
\centering
\includegraphics[width=\linewidth]{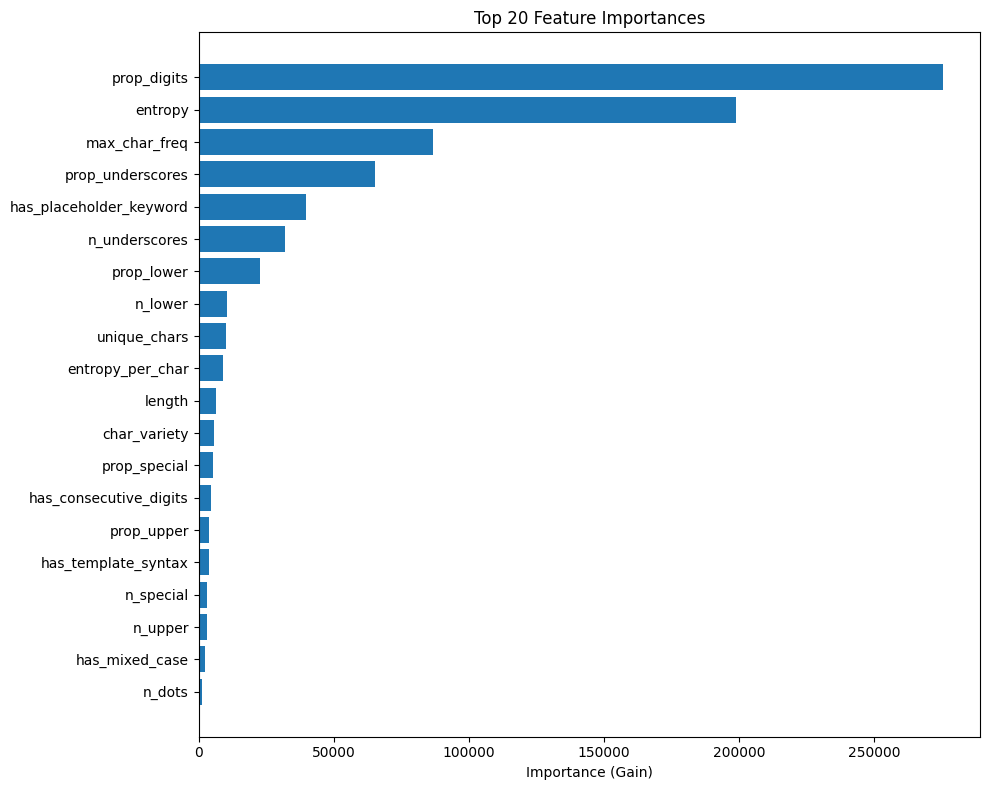}
\caption{Most important features in SECRET classification for LightGBM model}
\label{fig:secret_classification_features}
\end{figure}

\paragraph{Feature engineering and model training.}
Instead of fine-tuning a text classification model, we extract structured features from each string. Secrets are typically machine-generated and exhibit measurable statistical properties. We compute 43 features, and Figure \ref{fig:secret_classification_features} shows the top 20 most important features. The model found that the probability of a digit appearing in a string and the entropy of the string are the top 2 most important features in a real secret.

These features are assembled into a structured matrix and used to train a LightGBM classifier to predict whether a string is a real secret or a placeholder.

On a held-out synthetic test set of 15{,}000 samples ((7{,}500 samples from each class)), the model achieves near-perfect performance. The detailed breakdown is shown in Table~\ref{tab:secret-classifier}.

\begin{table}[h]
\centering
\caption{LightGBM performance on held-out synthetic test set (7{,}500 samples per class).}
\label{tab:secret-classifier}
\begin{tabular}{lccc}
\toprule
Class & Correct & Incorrect & Accuracy \\
\midrule
Placeholder & 7{,}482 & 18 & 99.76\% \\
Real Key & 7{,}474 & 26 & 99.65\% \\
\midrule
Overall & 14{,}956 & 44 & 99.71\% \\
\bottomrule
\end{tabular}
\end{table}

We note that high performance on synthetic data does not guaranty equivalent performance on real-world distributions.

\subsubsection{Model Classification}

We run the trained classifier on the 1{,}380 conversations identified via regex filtering. After removing strings classified as placeholders, 209 conversations remain.

Through manual inspection, 111 of the 209 conversations contain actual API secrets. Of these 111 conversations, 84 contain user-provided secrets. These 84 conversations correspond to 34 distinct API keys, as some users repeated the same key and prompt across multiple conversations.

For each detected secret, we compute a similarity-based confidence score between the model output and the user-provided text. When the confidence exceeds 0.5, we treat the secret as definitively user-provided.

Overall, this analysis confirms that real API keys and credentials appear in WildCode user prompts, even after removing placeholders and template artifacts.

We are unable to provide a list of our findings because some API keys are still in use.


\section{Re-prompting}
\label{sec:reprompt}
\subsection{methodology}
In the next phase of this study, we investigate whether commonly used prompt engineering strategies can lead to the generation of more secure code by large language models (LLMs). To this end, we conducted experiments across multiple LLMs and prompting strategies using prompts derived from the \textit{WildChat} dataset.

\begin{figure*}[t]
\centering
\includegraphics[width=\textwidth]{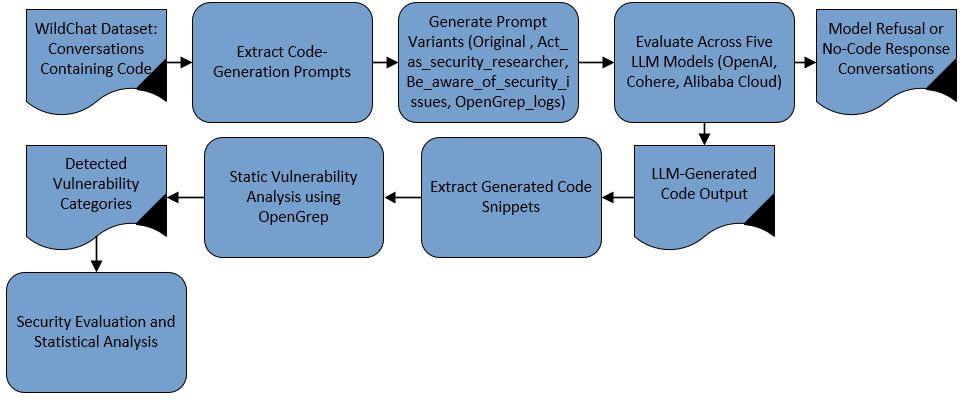}
\caption{Re-prompting methodology}
\label{fig:re-prompt}
\end{figure*}

Figure \ref{fig:re-prompt} illustrates the overall experimental pipeline used in this phase of our study. Code-generation prompts were first sampled from the WildChat dataset and filtered to retain conversations that produced executable code. Each selected prompt was then transformed into four prompting variants designed to evaluate the impact of different security-oriented instructions. These prompt variants were evaluated across five large language models from three providers: OpenAI, Cohere, and Alibaba Cloud. For each (prompt, model) combination, the generated code was collected and analyzed using the OpenGrep static analysis tool to identify potential vulnerabilities. Detected issues were categorized by vulnerability type and mapped to severity levels based on security impact. Finally, the resulting data were analyzed using multiple evaluation metrics, including vulnerability distribution analysis, severity-based risk scoring, and statistical tests, to assess how different prompting strategies influence the security characteristics of LLM-generated code.

From the Wildchat dataset, we selected prompts that resulted in \textbf{code generation tasks}. In total, \textbf{1126 prompts} that produced code outputs were identified and used in this experiment. These prompts served as the baseline user requests for evaluating the security properties of LLM-generated code.

To ensure experimental consistency with our previous analysis in Section \ref{sec:code analysis}, each model was provided with the same user's initial query, while we added different prefixes according to various prompting strategies. Specifically, we evaluated four prompting configurations:

\begin{itemize}
\item \textbf{Original:} The model receives the original user request with minimal formatting guidance:  
\textit{``Produce your solution in one code block'' + user\_prompt.}

\item \textbf{Act\_as\_security\_researcher:} The prompt instructs the model to assume the role of a security researcher when generating code:  
\textit{``Act as a security researcher and produce your solution in one code block'' + user\_prompt.}

\item \textbf{Be\_aware\_of\_security\_issues:} The prompt explicitly warns the model that the code contains security issues:  
\textit{``This code has a security issue and produce your solution in one code block'' + user\_prompt.}

\item \textbf{Opengrep\_log:} The prompt provides the model with the results of static security analysis obtained from OpenGrep:  

\textit{``This code has the following security issue and produce your solution in one code block:} +
\textit{\{LIST OF OPENGREP LOGS\}}
\textit{+ user\_prompt.''}
\end{itemize}

To evaluate the generality of prompting effects across different LLM architectures, we conducted experiments using five widely used models from three providers: OpenAI, Cohere, and Alibaba Cloud which are shown in table \ref{tab:reprompting_llms}. Additionally, Table \ref{tab:reprompting_languages} shows the sampled conversations from the WildCode dataset in which the conversation contained a vulnerable code reported by OpenGrep.

\begin{table}[H]
\centering
\begin{minipage}{0.48\textwidth}
\centering
\begin{tabularx}{\textwidth}{X X}
\hline
\textbf{Provider} & \textbf{Model} \\
\hline
OpenAI & gpt\_oss\_120b \\
OpenAI & gpt\_oss\_20b \\
Cohere & command\_r7b \\
Cohere & command\_a \\
Alibaba Cloud & qwen3-30b-a3b-instruct-2507 \\
\hline
\end{tabularx}
\caption{Evaluated large language models and their providers.}
\label{tab:reprompting_llms}
\end{minipage}
\hfill
\begin{minipage}{0.48\textwidth}
\centering
\begin{tabularx}{\textwidth}{X c}
\hline
\textbf{Language} & \textbf{Prompts} \\
\hline
Python & 248  \\
C & 238 \\
Java & 235 \\
JavaScript & 237  \\
PHP & 49 \\
C\# & 119 \\
\hline
\end{tabularx}
\caption{Number of conversations retained per programming language.}
\label{tab:reprompting_languages}
\end{minipage}
\end{table}

\begin{figure}[H]
\centering
\includegraphics[width=0.8\linewidth]{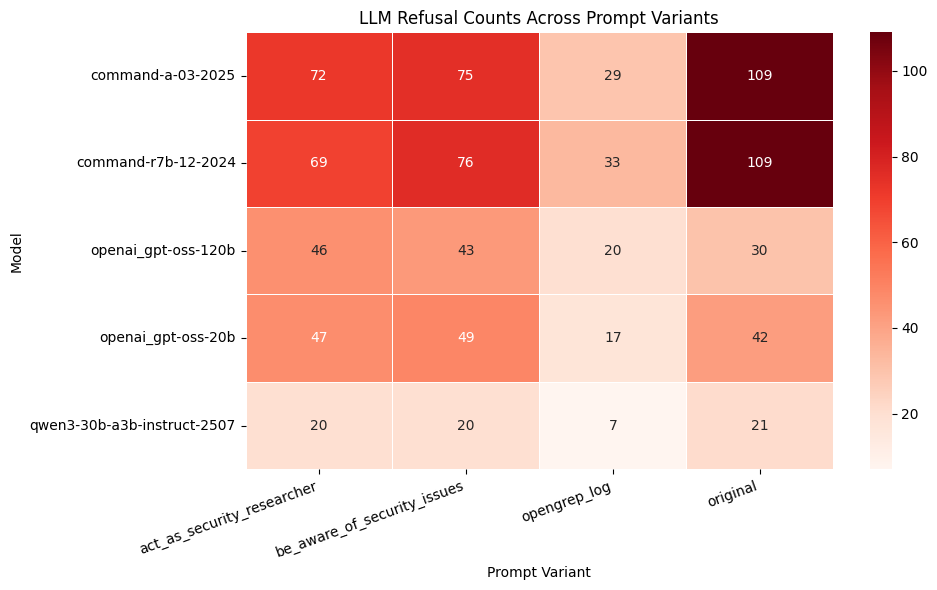}
\caption{Heatmap showing the number of refusals across models and prompt variants. Darker colors indicate higher refusal counts.}
\label{fig:refusal_heatmap}
\end{figure}

In some cases, LLMs opted not to generate code, even though the earlier model (from the Wildchat dataset) had done so. Figure \ref{fig:refusal_heatmap} reports the number of conversations, out of 1126 prompts, where models refused to provide code. Refusal reasons varied according to the model: OpenAI models often used similar wording, suggesting fine-tuning, while Cohere models typically refused when prompts involved malicious code or ethical risks. Overall, Cohere models were more reluctant to provide code, indicating stronger ethical alignment. Out of all models, Qwen from Alibaba had the least refusals and would generate code snippets given a wide range of prompts. The refusals act as a limitation to our statistics, as they decrease the number of conversations we can testing for vulnerabilities.

For each (prompt, model) combination, the generated code was collected and analyzed using a vulnerability detection pipeline. To evaluate the security posture beyond raw vulnerability counts, each vulnerability category was manually assigned a severity level: \textit{high}, \textit{medium}, or \textit{low}. The classification is informed by widely recognized security resources, including the OWASP Top 10, common CVSS severity ranges, exploitability characteristics, and expected impacts on data confidentiality, integrity, and availability. Categories associated with remote code execution, authentication bypass, data exfiltration, or strong integrity compromise were classified as \textit{high}. Context-dependent issues (e.g., cross-site scripting or information disclosure) were classified as \textit{medium}, while configuration-level or lower exploitability issues were classified as \textit{low}. For consistency in analysis, severity levels were mapped to numeric weights: High = 5, Medium = 3, and Low = 1. The specific mapping of vulnerability categories to severity levels is summarized below.

\begin{itemize}
\item \textbf{High severity:} command injection, buffer overflow, deserialization, authentication and authorization bypass (authn\_authz), path traversal, server-side request forgery (SSRF), hardcoded secrets, XML external entity (XXE), SQL injection (SQLi), insecure cryptographic implementation
\item \textbf{Medium severity:} cross-site scripting (XSS), cross-site request forgery (CSRF), format string vulnerability, open redirect, information leakage
\item \textbf{Low severity:} TLS/SSL configuration issues, insecure random number generation, denial-of-service (DoS), other minor security issues
\end{itemize}

\subsection{Evaluation Metrics}

To quantitatively evaluate the security implications of different prompting strategies, we employed two complementary analysis approaches: a severity-weighted security risk metric and distributional divergence analysis. Together, these metrics allow us to assess both the \textit{severity burden} of vulnerabilities and the \textit{structural changes} in vulnerability distributions across prompts and models.

\paragraph{Normalized Security Risk Score (NSRS).}
To measure the overall severity of vulnerabilities produced by each model--prompt combination, we compute a \textit{Normalized Security Risk Score (NSRS)}. For each generated code sample, vulnerabilities detected by the analysis pipeline are assigned severity weights based on their classified severity levels. In our analysis, high-, medium-, and low-severity vulnerabilities are assigned numeric weights of 5, 3, and 1, respectively. The weighted severity score is calculated by summing the assigned weights of all detected vulnerabilities. To ensure comparability across prompts that may produce different numbers of vulnerabilities, the total severity score is normalized by the total number of detected vulnerabilities:

\begin{equation}
NSRS = \frac{\sum_{i=1}^{n} w_i}{n}
\end{equation}

where $w_i$ represents the severity weight of vulnerability $i$, and $n$ represents the total number of detected vulnerabilities. This normalization captures the \textit{average vulnerability severity} produced by a given prompt and model. Lower NSRS values indicate that generated code tends to contain vulnerabilities with lower severity, suggesting improved security characteristics.

\paragraph{Vulnerability Distribution Analysis.}
While the NSRS metric captures severity trends, it does not capture how prompting strategies influence the \textit{distribution of vulnerability categories}. To analyze these structural changes, we compute the Kullback--Leibler (KL) divergence \cite{kullback1951information} between vulnerability distributions generated under different prompts and the baseline prompt distribution.

For each programming language, we construct contingency tables where rows represent vulnerability categories and columns represent prompt--model combinations. Raw counts are converted to probability distributions to allow fair comparisons across prompts with different vulnerability counts. KL divergence is then computed as:

\begin{equation}
D_{KL}(P \parallel Q) =
\sum_i P(i)\log\frac{P(i)}{Q(i)}
\end{equation}

where $P(i)$ represents the probability of vulnerability category $i$ under a given prompt configuration, and $Q(i)$ represents the probability under the baseline prompt. A divergence value of zero indicates identical vulnerability distributions, while larger values indicate greater deviation from the baseline behavior.

\paragraph{Statistical Significance Testing.}
To determine whether the  differences observed in vulnerability distributions and severity across prompting strategies are statistically meaningful, we conducted a set of statistical significance tests. First, a \textit{Chi-square test of independence} was applied to contingency tables constructed from vulnerability categories and prompt types. This test evaluates whether the distribution of vulnerability categories is independent of the prompting strategy or whether certain prompts are associated with specific types of vulnerabilities. In other words, the test determines whether the observed frequencies of vulnerability categories across prompts differ significantly from what would be expected if prompts had no effect on vulnerability generation. To further quantify the strength of this association, we computed \textit{Cramér’s V}, which provides a normalized effect size for the Chi-square statistic. This metric allows us to assess not only whether the relationship is statistically significant but also how strong the relationship is between prompt type and vulnerability category.

In addition to analyzing differences in vulnerability distributions, we also examined whether the \textit{severity of vulnerabilities} generated by different prompting strategies differs significantly. Because severity scores are ordinal and may not follow a normal distribution, we employed the \textit{Kruskal–Wallis test}, a non-parametric alternative to one-way ANOVA \cite{kruskal1952use}. This test evaluates whether the median severity scores differ across multiple prompting strategies. When the Kruskal–Wallis test indicated statistically significant differences, we performed post-hoc pairwise comparisons using \textit{Dunn’s test} with Bonferroni correction. This post-hoc analysis identifies which specific prompt pairs exhibit statistically significant differences in vulnerability severity while controlling for the increased risk of Type I error due to multiple comparisons.

\subsection{Results and Findings}

\begin{figure}[t]
\centering
\includegraphics[width=\linewidth]{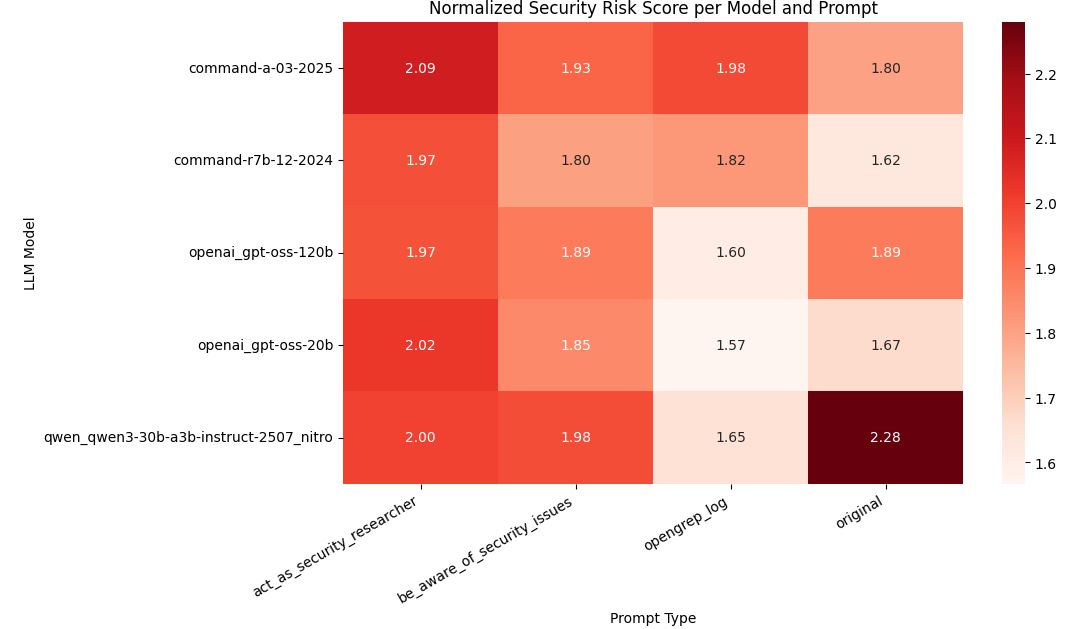}
\caption{Prompt--Model Security Landscape based on the Normalized Security Risk Score (NSRS).}
\label{fig:nsrs}
\end{figure}

\paragraph{Normalized Security Risk Score (NSRS).}
Figure~\ref{fig:nsrs} presents the normalized security risk score for each LLM model and prompt type. Each cell represents the average severity of vulnerabilities generated by a given LLM model under a specific prompting strategy. Lower values indicate safer outputs, meaning the generated code contains vulnerabilities with lower average severity, while higher scores indicate that the generated code tends to contain vulnerabilities with greater severity.

Overall, the results show that prompt engineering can influence the security characteristics of generated code, although the magnitude of this effect varies across models. Among the evaluated prompting strategies, the \textit{opengrep\_log} prompt consistently produces the lowest risk scores for most models. This suggests that incorporating static analysis feedback into the prompt helps guide the models toward generating code with less severe vulnerabilities. For example, the \textit{command-r7b-12-2024} model achieves its lowest score (1.62) under the \textit{opengrep\_log} prompt, while the \textit{openai\_gpt-oss-20b} model also performs best with this prompt (1.57). Similarly, the \textit{command-a-03-2025} model shows its lowest risk score (1.80) under the same configuration, suggesting that the inclusion of OpenGrep analysis improves security outcomes across several models.

The \textit{be\_aware\_of\_security\_issues} prompt generally produces moderate improvements, with risk scores slightly lower than those obtained using the \textit{act\_as\_\linebreak researcher} prompt. This indicates that explicitly reminding the model about security concerns can reduce vulnerability severity in some cases, although the effect is less consistent than when static analysis feedback is included.

In contrast, the \textit{act\_as\_security\_researcher} prompt does not consistently improve security outcomes. Several models still produce relatively high risk scores under this configuration. For instance, the \textit{command-a-03-2025} model reaches a risk score of 2.09, which is among the highest values observed. This suggests that simply instructing the model to adopt the role of a security researcher does not reliably guide it toward producing safer code.

Finally, the \textit{original} prompt, which repeats the user request without additional security guidance, often produces higher risk scores compared to the other prompting strategies. This is particularly evident for the \textit{qwen3-30b-a3b-instruct-2507} model, which reaches the highest observed score (2.28) under the original prompt. This indicates that without explicit security guidance, some models tend to generate code containing more severe vulnerabilities.

Overall, these results suggest that prompt augmentation can meaningfully influence the security properties of LLM-generated code, with prompts that incorporate external security analysis signals—such as OpenGrep logs—providing the most consistent reductions in vulnerability severity across models.

\begin{figure}[t]
\centering
\includegraphics[width=\linewidth]{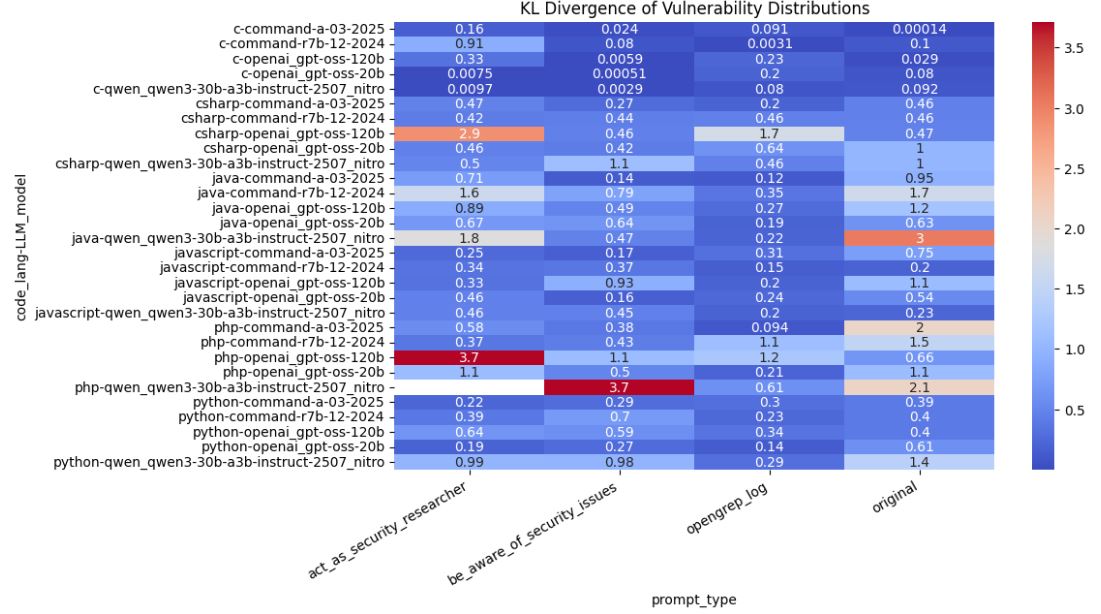}
\caption{KL divergence of vulnerability category distributions across prompting strategies and programming languages.}
\label{fig:kl_divergence}
\end{figure}

\paragraph{Vulnerability Distribution Analysis (KL Divergence).}
To analyze how prompting strategies influence the distribution of vulnerability categories, we computed the Kullback--Leibler (KL) divergence between vulnerability distributions produced under different prompts and those observed under the baseline prompt.

the KL divergence analysis reveals that prompt engineering can significantly influence vulnerability distributions, although the magnitude of this effect varies across programming languages and model configurations. As  is shown in Figure \ref{fig:kl_divergence}, most model–prompt combinations exhibit relatively low divergence values ($<1$), indicating that many prompts produce vulnerability distributions similar to those observed under the baseline WildChat prompt. This suggests that simply repeating the original user prompt with different model versions often results in vulnerability patterns that closely resemble the baseline behavior.

However, several configurations exhibit substantially higher divergence values, indicating that prompt augmentation can significantly shift vulnerability distributions. In particular, PHP configurations show the largest divergence values, especially for the \textit{act\_as\_security\_researcher} and \textit{be\_aware\_of \_security\_issues} prompts, where divergence values exceed 3.0. This suggests that explicitly instructing models to reason about security can substantially alter the vulnerability patterns produced in generated code.

Moderate divergence values are also observed for Java and C\# configurations, indicating that prompt-based guidance can influence the types of vulnerabilities introduced or mitigated in these languages. In contrast, Python and JavaScript configurations exhibit relatively low divergence values, suggesting that vulnerability distributions remain more stable across prompt variations.

The \textit{opengrep\_log} prompt typically produces moderate divergence values, indicating that incorporating static analysis feedback can alter vulnerability distributions, although the effect varies across languages and models. In some cases, the static analysis context appears to guide the model toward different vulnerability categories by highlighting previously detected issues.

Overall, the KL divergence analysis demonstrates that prompt modifications can influence the vulnerability characteristics of LLM-generated code, but the extent of this influence depends on both the programming language and the underlying LLM architecture.

\subsection{Statistical Significance Testing}

To evaluate whether the observed differences across prompting strategies are statistically meaningful, we conducted several statistical tests examining both vulnerability distributions and vulnerability severity.

\textbf{Vulnerability Distribution Differences.}
To determine whether prompting strategies influence the types of vulnerabilities generated by LLMs, we performed a \textit{Chi-square test of independence} on contingency tables constructed from vulnerability categories and prompt types. The results show a statistically significant association between prompt strategy and vulnerability category ($\chi^2 = 239.37$, $p-value < 0.001$). This indicates that the distribution of vulnerability categories is not independent of the prompt configuration and that prompting strategies significantly influence the types of vulnerabilities produced by LLM-generated code.

To quantify the magnitude of this relationship, we computed \textit{Cramér’s V}, which resulted in $V = 0.1166$. According to common interpretation thresholds, this corresponds to a small effect size. This suggests that although prompting strategies significantly affect vulnerability distributions, the overall magnitude of this effect remains modest. This observation is consistent with the KL divergence analysis, where most prompt–model configurations exhibit relatively low divergence values (below 1), indicating that vulnerability distributions remain broadly similar to the baseline prompt despite observable shifts in specific cases.

\textbf{Vulnerability Severity Differences.}
To evaluate whether prompting strategies influence the severity of generated vulnerabilities, we performed a \textit{Kruskal– Wallis test} on the severity scores derived from the assigned vulnerability weights. The test results indicate a statistically significant difference in vulnerability severity across prompting strategies ($H = 21.79$, $p-value < 0.001$). This finding confirms that prompt design affects not only the types of vulnerabilities produced by LLMs but also their overall severity profile.

To identify which prompting strategies differ significantly, we conducted post-hoc pairwise comparisons using \textit{Dunn’s test} with Bonferroni correction. The results reveal significant differences between the \textit{act\_as\_security\_researcher} prompt and both the \textit{opengrep\_log} prompt ($p-value = 0.00007$) and the \textit{original} prompt ($p-value = 0.035$). In contrast, other prompt pairs do not exhibit statistically significant differences. These findings indicate that the \textit{act\_as\_security\_ researcher} prompt behaves differently from other prompting strategies with respect to vulnerability severity.

Taken together, the statistical tests support the observations from the NSRS and KL divergence analyses. Prompt engineering significantly affects both vulnerability distributions and severity levels in generated code. However, the relatively small effect size observed in the Cramér’s V analysis suggests that while prompting strategies can influence security outcomes, they do not fundamentally alter the overall vulnerability behavior of LLM-generated code across all cases.

\section{User Intent}
\label{sec:user}
In this section, we investigate how users engage with ChatGPT in code-related conversations, focusing on four complementary dimensions: (i) ChatGPT’s implicit programming language preferences when users do not specify a language, (ii) the intentions that underlie user queries, (iii) the relationship between user intent and the total length and depth of conversations, and (iv) the extent to which users address security concerns when interacting with the code generated by ChatGPT. By analyzing these aspects in the conversations related to code in Wildchat, our goal is to uncover patterns in user behaviour, identify gaps in security awareness, and better understand how ChatGPT mediates coding practices across multi-turn dialogues.

To better understand the intentions of users and their interpretation of coding issues, we restrict our analysis to queries formulated in English. These queries span a wide spectrum of concerns, including code understanding, bug identification and resolution, debugging practices, security vulnerabilities, and performance optimization. For this purpose, we have used the WildCode\_EN dataset, which contains the English conversations from the WildChat dataset that also contain code, as explained in Section \ref{sec:dataset}.

\subsection{ChatGPT programming language preference}
Our first analysis investigates ChatGPT’s default programming language choices when users omit specifying a language in their requests. The goal is to understand the implicit defaults of ChatGPT in code generation, as these defaults shape the coding environment presented to users, influence the accessibility of generated solutions, and may reveal underlying model biases toward certain languages (e.g., Python). For this analysis, we use the \textit{WildCode} dataset. The programming languages of the code snippets in these conversations were determined based on the language labels assigned to the code snippets in \textit{WildCode}, as described in Section~\ref{sec:dataset}. We extracted all conversations in which the user did not specify any programming language in their \textit{initial query}, while ChatGPT's response included a code snippet.

\begin{figure}[ht]
    \centering
    \begin{subfigure}[b]{0.55\textwidth}
        \centering
        \includegraphics[width=\textwidth]{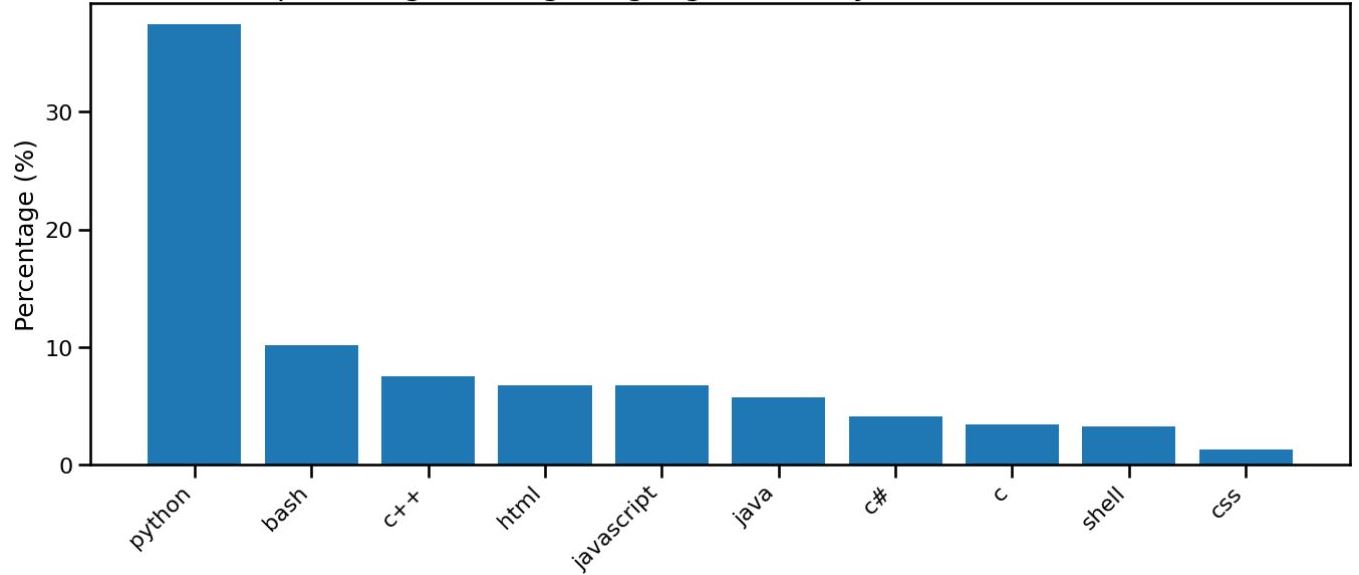}
        \caption{Top 10 programming languages used by ChatGPT in initial code generation.}
        \label{fig:init_code_languages}
    \end{subfigure}
    \hfill
    \begin{subfigure}[b]{0.43\textwidth}
        \centering
        \includegraphics[width=\textwidth]{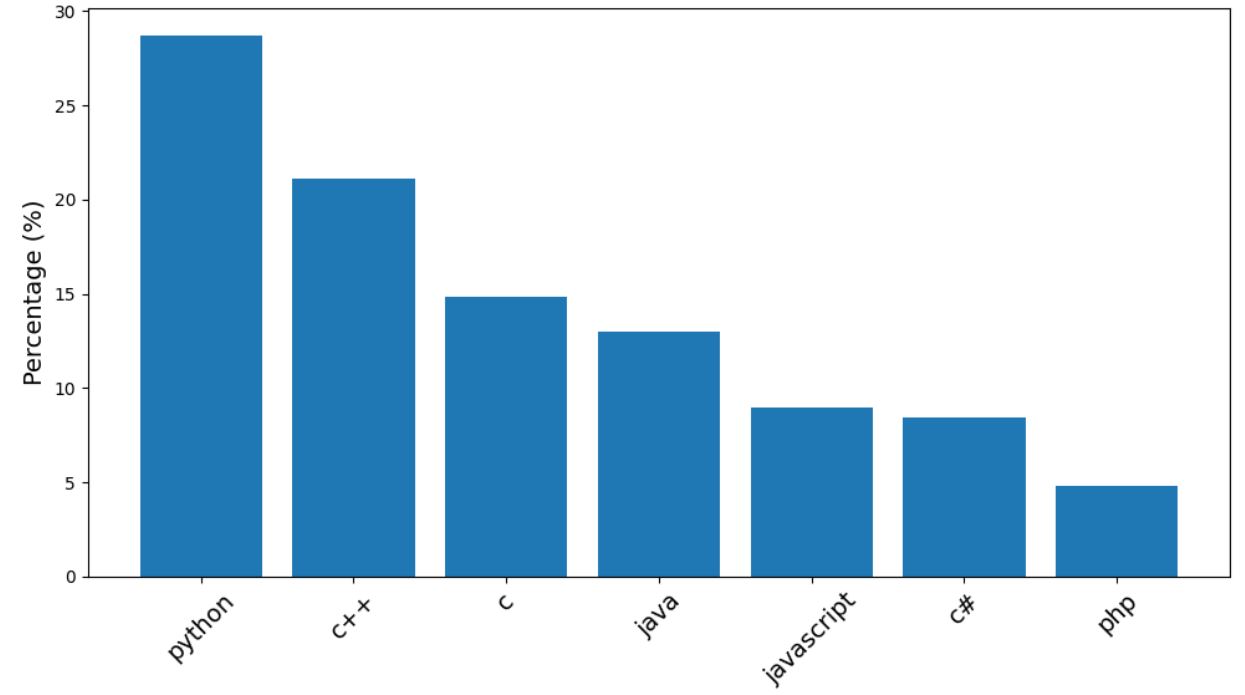}
        \caption{Languages newly requested by users in follow-up queries.}
        \label{fig:new_languages_followup}
    \end{subfigure}
    \caption{Comparison of programming languages in initial code generation vs. user-requested follow-up.}
    \label{fig:lang_comparison}
\end{figure}

Figure~\ref{fig:init_code_languages} illustrates the distribution of programming languages in the code snippets generated by ChatGPT during conversations where the user did not specify any programming language. As can be seen, ChatGPT exhibits a marked preference for writing Python code, which accounts for more than one-third of all generated code, followed by Bash, C++, HTML, and JavaScript. This suggests a strong default preference or widespread applicability of Python in initial coding tasks. A sample conversation is provided in the project's Github.

Figure~\ref{fig:new_languages_followup} shows the distribution of the programming languages requested by the users in \textit{follow-up queries}. While Python remains the most frequently mentioned language; C++, C, and Java are also commonly requested when the code initially generated by ChatGPT was in a different language. This pattern indicates that users often shift programming languages during multi-turn interactions, possibly due to evolving task requirements or preferences.


\vspace{-15pt}

\subsection{Users' intentions in code related queries}

In the next stage of our analysis, we investigate users’ intentions in code-related queries, with particular emphasis on follow-up messages. Understanding these intentions is crucial for characterizing how users engage with ChatGPT beyond their initial requests, as follow-up queries often reveal deeper goals such as clarifying outputs, fixing errors, or adapting code to new requirements. 

To understand users' intentions behind their code-related queries, we defined a set of categories that represent common types of coding requests. The definitions of these categories are provided in the paper's github repository. For classification, we first removed code snippets from user messages, retaining only the natural language text. We then applied zero-shot classification using the \texttt{bart-large-mnli}\footnote{\url{https://huggingface.co/facebook/bart-large-mnli}} model, leveraging the category definitions and example keywords as candidate labels for intent detection. The model produced a probability distribution over the predefined categories for each query. To ensure fairness when multiple categories received nearly identical confidence levels, the probabilities were rounded to two decimal places, and all categories that shared the maximum rounded probability were selected. This tie-aware approach captures cases where the model could not clearly differentiate between categories, instead of arbitrarily selecting only one. Using this methodology, we derived three types of labels for each conversation: (i) the initial category based on the user’s first query, (ii) the primary follow-up category corresponding to the second query in the query sequence in a conversation with ChatGPT, and (iii) aggregated follow-up categories in subsequent queries except for the initial.

\begin{figure}[ht]
    \centering
    \includegraphics[width=\linewidth]{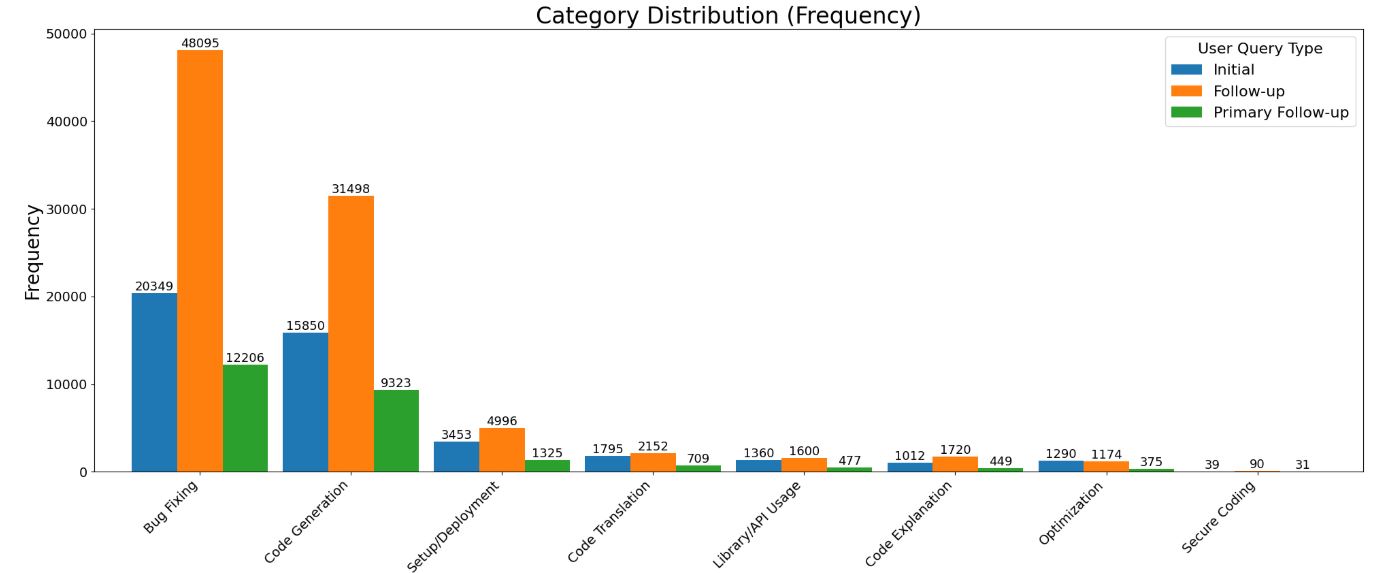}
    \caption{
        Predicted category distribution for user queries. 
    }
    \label{fig:conversation_category}
\end{figure}

Figure~\ref{fig:conversation_category} presents the distribution of user intents across these three contexts. \textit{Bug Fixing} and \textit{Code Generation} emerge as the most frequent categories, reflecting that practical coding support is the predominant concern at the start of user interactions with ChatGPT. Similarly, in follow-up queries, \textit{Bug Fixing}, \textit{Code Generation}, and \textit{Setup/Deployment} remain dominant. In contrast, categories such as \textit{Secure Coding} and \textit{Optimization} appear much less frequently, suggesting that these considerations are less commonly prioritized during user–assistant exchanges.

We also examined the relationship between users' initial query categories and the categories of their follow-up messages. Table~\ref{tab:followup_by_initial} presents the most common follow-up categories associated with each initial query type. A clear pattern emerges: \textit{Bug Fixing} and \textit{Code Generation} dominate the follow-up space across nearly all initial categories. This indicates that, regardless of the original intent, users frequently transition toward refining existing code or requesting new code during their interactions. Notably, \textit{Code Generation} appears frequently both as an initial intent and as a follow-up category, highlighting the iterative nature of coding workflows with ChatGPT.

\begin{table}[ht]
\centering
\caption{Most Common Follow-up Categories by Initial Query Type}
\label{tab:followup_by_initial}
\begin{tabular}{l|l}
\toprule
\textbf{Initial Category (Count)} & \textbf{Most Common Follow-up Category} \\
\midrule
Bug Fixing (8017)         & Bug Fixing (5034), Code Generation (2983) \\
Code Explanation (510)    & Bug Fixing (292), Code Generation (218) \\
Code Generation (8017)    & Code Generation (4126), Bug Fixing (3891) \\
Code Translation (783)    & Bug Fixing (456), Code Generation (327) \\
Library/API Usage (593)   & Bug Fixing (360), Code Generation (233) \\
Optimization (383)        & Bug Fixing (312), Secure Coding (71) \\
Secure Coding (21)        & Code Explanation (11), Bug Fixing (10) \\
Setup/Deployment (1535)   & Bug Fixing (949), Code Generation (586) \\
\bottomrule
\end{tabular}
\end{table}

Interestingly, when \textit{Secure Coding} is the initial focus, it rarely results in continued security-related discussions; suggesting a gap in users' sustained engagement with secure development practices. Conversely, follow-up queries to \textit{Optimization} tasks occasionally involve \textit{Secure Coding}, implying that some users perceive a connection between performance and security. To statistically assess these patterns, we applied the chi-square test of independence. The test results revealed no significant association between initial and follow-up categories, indicating that despite the apparent trends, such as the dominance of \textit{Bug Fixing} and \textit{Code Generation}, these follow-up intent categories are not strongly dependent on the user's initial query category.


\subsection{Impact of Query Category on Conversation Length}

We next examine how the category of a user’s initial or follow-up query influences the overall length of the conversation with ChatGPT. This analysis provides insight into whether certain types of coding requests, such as bug fixing, code explanation, or secure coding, tend to generate more extended interactions, while others can be resolved more quickly. Here, \textbf{conversation length} is measured as the number of user–ChatGPT query–response pairs within a conversation. By linking query categories to length, we aim to better understand the dynamics of multi-turn dialogs and the factors that drive longer or shorter exchanges.

\begin{figure}[ht]
    \centering
    \includegraphics[width=1\textwidth]{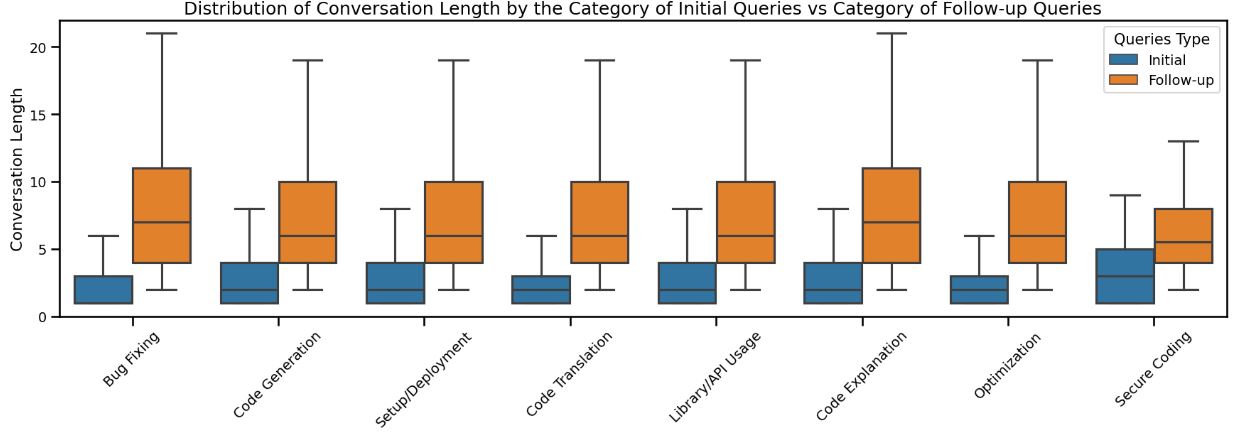}
    \caption{Distribution of conversation length grouped by initial and followup message category.}
    \label{fig:conversation_turns}
\end{figure}

Figure~\ref{fig:conversation_turns} illustrates the distribution of conversation lengths, measured by the total number of messages exchanged per conversation, grouped by initial and follow-up categories, respectively. These visualizations provide insight into how the nature of a user’s request influences the depth and complexity of the ensuing interaction. As shown in Figure~\ref{fig:conversation_turns}, conversations that begin with \textit{Secure Coding} tend to have the highest median and widest range in message counts, suggesting that security-focused topics often prompt more extensive discussions. In contrast, queries related to \textit{Bug Fixing} and \textit{Code Translation} are typically resolved in shorter conversations, indicating these are more concise or well-scoped tasks.

Figure~\ref{fig:conversation_turns} shows that when the follow-up category is \textit{Code Explanation} or \textit{Bug Fixing}, conversations tend to be longer, potentially due to the need for iterative clarification or detailed reasoning. Meanwhile, follow-up requests involving \textit{Secure Coding} exhibit fewer messages, implying that even when security is addressed later in a conversation, users do not typically engage in extended dialogue on that topic.

Overall, the number of messages exchanged appears to reflect the perceived complexity or ambiguity of the task, with explanatory and security-related queries tending to foster longer, more in-depth interactions.

\subsection{Security Awareness in Code-Related Conversations}


We now turn our attention to the extent to which users address security concerns in their interactions with ChatGPT-generated code. While prior analyses have shown that users frequently focus on practical tasks such as code generation and bug fixing, it remains unclear how often security considerations enter these conversations. By examining both initial query intents and follow-up queries, we aim to identify whether users explicitly engage with secure coding practices, how often security arises in multi-turn dialogues, and whether it is sustained throughout the interaction. This analysis provides critical insight into the role of security awareness in AI-assisted coding workflows.

As mentioned in Section \ref{sec:code analysis}, we analyzed code snippets generated by ChatGPT using \textit{OpenGrep} to identify instances containing errors.  Out of 48,391 conversations that include code, the code generated by ChatGPT in 1,562 conversations were flagged by OpenGrep as having at least one error. Among these, 1,214 conversations were conducted in English, which we used as the basis for our intent analysis. 

\begin{figure}[th] 
\centering \includegraphics[width=0.9\linewidth]{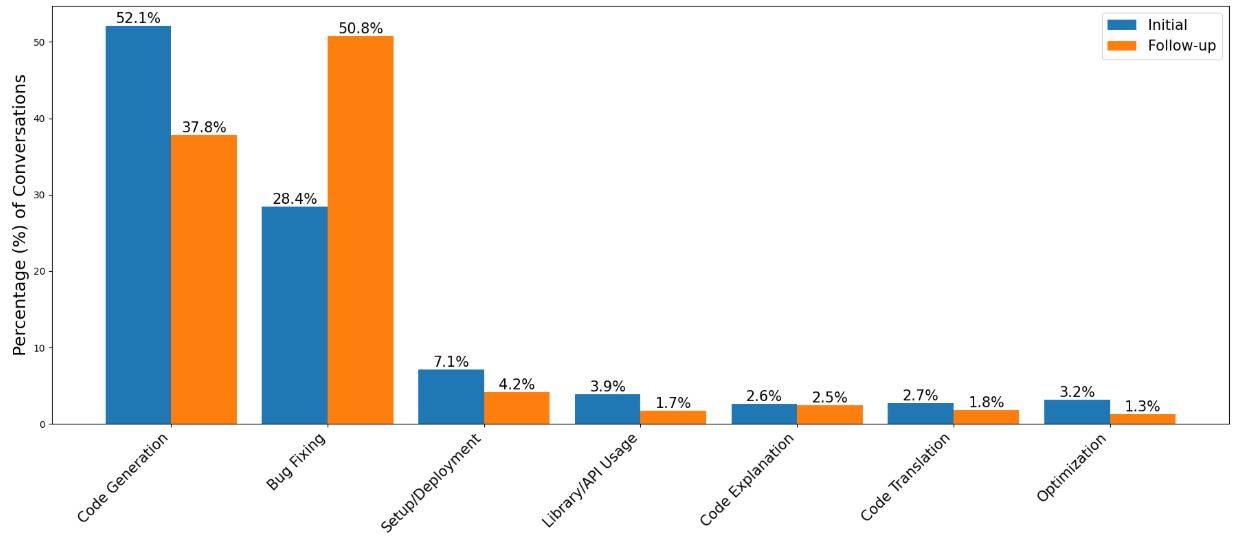} \caption{Predicted Category for User Query in Conversations with Buggy Codes} \label{fig:category_buggy} 
\end{figure}
\vspace{-5pt}

Figure \ref{fig:category_buggy} presents the distribution of the predicted intent category for initial and follow-up queries from the user in these buggy code conversations. A key observation is that while \textit{Code Generation} and \textit{Bug Fixing} dominate both initial queries and follow-up actions, \textit{Secure Coding} is extremely rare, with only 6 instances in all follow-ups. This suggests a noticeable gap in user emphasis on security-related intents, highlighting the need for increased security awareness and better integration of secure coding practices in AI-assisted development. 

\subsection{User Discussion on Hallucinated Modules}
As discussed in Section \ref{sec:hallucinations}, ChatGPT generated non-existent (hallucinated) modules in some of the produced code snippets. We manually examined the list of conversations\footnote{Available on our GitHub repository: \url{https://github.com/regularpooria/WildCode}} where these modules were mentioned, and the conversations were conducted in English. In none of these cases did the users suspect that the module might be fake; instead, they typically continued asking about the errors they encountered. Likewise, ChatGPT did not self-correct or indicate that the issue could originate from a non-existent module. An example of such a conversation involving a hallucinated module is presented in our GitHub.

\subsection{Refusal Circumvention and Safety Constraint Bypass}

\label{sec:Refusal}
We then turned our attention to identifying reasons why the model would refuse to generate code, if asked to do so by the user.  We extracted conversations in which the prompt had led to creation of code in the Wildchat dataset, but the model refused to generate code in our replication experiment.  Specifically, we implemented a regular-expression (regex) function to identify responses in which the model declined to provide assistance. The refusal patterns captured common linguistic indicators, including apologetic expressions (e.g., ``I'm sorry''), explicit inability statements (e.g., ``I cannot,'' ``I can't''), and references to illegality, malware, policy violations, or security-related constraints. Using this automated procedure, we extracted conversations in which the model exhibited refusal behaviour and categorized them according to the matched regex pattern.

The  categories  identified reflect distinct refusal rationales observed in practice:

\begin{itemize}
    \item \textbf{cannot\_statement}: Captures explicit denials of capability or permission (e.g., ``I cannot help with that''). These refusals may arise from safety constraints, scope or feasibility limitations, incomplete specifications, or excessive task size. Examples include requests to generate extremely large-scale systems without explanation (e.g., ``write a complete enterprise banking system with no comments''), reproduce full proprietary applications, or construct extensive visualization pipelines without context.

    \item \textbf{apology}: Includes responses framed as polite refusals, typically using softened language (e.g., ``I'm sorry, but I can't assist with that request''). These often occur in response to requests involving harmful code, privacy invasion, academic misconduct, or other restricted content, even when policy is not explicitly cited.

    \item \textbf{mentions\_illegal}: Corresponds to refusals triggered by requests involving unlawful activity or intellectual property violations. Examples include hacking accounts, bypassing authentication mechanisms, conducting fraud, generating pirated software, reproducing proprietary source code, or requesting copyrighted application implementations.

    \item \textbf{mentions\_malware}: Includes refusals explicitly referencing malicious software development. Typical examples involve ransomware, spyware, keyloggers, botnets, remote access trojans, exploitation scripts, or denial-of-service tools. These refusals are generally justified on harm-prevention grounds.

    \item \textbf{policy\_violation}: Captures responses that directly reference violations of usage policies or service terms (e.g., ``this request violates policy''). These commonly involve harmful automation, credential harvesting, large-scale scraping, or operational misuse of systems.

    \item \textbf{security\_reason}: Reflects refusals justified by safety or security considerations, even when illegality is not explicitly mentioned. Examples include requests to bypass safeguards, disable authentication checks, remove encryption, expose credentials, or modify systems in ways that introduce vulnerabilities.
\end{itemize}

The primary objective was to characterize the reasons for model refusal and to evaluate whether subsequent user follow-up prompts were able to induce code generation after an initial refusal. Rather than labelling such cases as prompt injection, we interpret them more conservatively as instances of \emph{refusal circumvention} or \emph{safety constraint bypass}, where iterative or adaptive prompting resulted in eventual compliance. 

For categories with relatively small sample sizes (i.e., \textit{mentions\_illegal}, \textit{security\_reason}, \textit{mentions\_malware}, and \textit{policy\_violation}), we conducted a full manual review of all conversations. Two independent researchers examined each case to determine whether the user successfully induced code generation following the initial refusal.

For the larger categories (\textit{cannot\_statement} and \textit{apology}), we employed a two-stage filtering procedure. In the first stage, an OpenAI model (\textit{gpt-4o-mini}) was used to summarize conversations and identify candidate cases in which the model initially refused but subsequently generated code following a user follow-up prompt. This automated screening identified 99 out of 1,085 conversations in the \textit{cannot\_statement} category and 29 out of 533 conversations in the \textit{apology} category as potential refusal-circumvention instances. In the second stage, all conversations selected by the model during automated screening were manually reviewed by two independent researchers to verify whether a genuine refusal-circumvention event had occurred, defined as code generation following an explicit initial refusal. 

Table~\ref{tab:init_refusal_categories} presents the  refusal categories we identified and the corresponding number of conversations in each category. The hash identifiers of the validated refusal-circumvention conversations are publicly released in our GitHub repository to serve as a ground-truth benchmark dataset for evaluating model robustness against safety-constraint bypass in code-generation contexts.

\begin{table}[ht]
\centering
\caption{Initial Refusal Categories and Refusal-Circumvention Outcomes}
\label{tab:init_refusal_categories}
\begin{tabular}{l c c}
\hline
\textbf{Refusal Category   } & \textbf{Number of    } & \textbf{Refusal-Circumvention } \\
\textbf{  } & \textbf{ Conversations   } & \textbf{ Conversations} \\
\hline
cannot\_statement   & 1085 & 26 \\
apology             & 533  & 25 \\
mentions\_illegal   & 152  &40 \\
security\_reason    & 41   & 0 \\
mentions\_malware   & 34   & 7  \\
policy\_violation   & 14   & 14 \\
\hline
\end{tabular}
\end{table}

\section{Discussion and Threat to Validity}
\label{sec: Diss}
Overall, our study suggests that ChatGPT-generated code is often of low quality, particularly in terms of security. While newer models or coding-specific tools (e.g., GitHub Copilot) may yield better results, this remains to be verified.

Our vulnerability detection likely overcounts some issues (e.g., vulnerable regex or deserialization used in safe contexts), while undercounting others, particularly memory issues in C/C++, due to limited pattern coverage. The use of lightweight regex-based analysis ensures scalability, but it may miss context-sensitive flaws. Although we manually verified the detected cases, some insecure snippets may remain undetected, and our reported rates should be considered lower bounds.

Our user intent analysis excludes non-English conversations to ensure consistency, which may introduce bias by omitting multilingual interaction patterns. In contrast, our security analysis includes code from all languages, potentially affecting comparability. Additionally, zero-shot classification, while scalable, may miss nuanced intents, suggesting that future work explore more tailored models.

The WildChat dataset omits interactions involving code shared as images, which may underestimate the diversity of code-related queries. Finally, while a few users explicitly request vulnerable code (e.g., buffer overflow examples), such cases are rare and unlikely to impact our overall findings. Notably, ChatGPT often refuses to generate attack code but readily provides vulnerable code—a curious ethical asymmetry.

\section{Conclusion}
\label{sec: Conclu}

In this paper, we present the first large-scale empirical study of code generated by ChatGPT based on real-world user interactions drawn from the WildChat dataset. Unlike previous research that relies on synthetic prompts or benchmark datasets, our analysis leverages 82,843 authentic conversations in which ChatGPT generates code, allowing us to assess both the quality of the code and the intentions of the users who request it.

Our findings highlight several concerning trends. First, code quality, particularly in terms of security, remains a significant issue. Second, our analysis of user intent shows that security is rarely prioritized in user queries. Even when users encounter buggy or vulnerable code, they seldom raise security concerns or request secure alternatives.Together, these contributions underscore the urgent need for security-aware LLMs, better user prompting strategies, and proactive safeguards within generative coding tools.



\section{Declaration}
\subsection{Funding}
This work received in-kind support from Cohere in the form of complimentary API usage credits (approximately USD\$1,000), which were used to conduct experiments evaluating prompting strategies and their impact on the security characteristics of LLM-generated code. The funder had no role in the design of the study; in the collection, analysis, or interpretation of data; in the writing of the manuscript; or in the decision to publish the results.
\subsection{Acknowledgements}
The authors would like to thank the Centre for Applied Artificial Intelligence (CAAI) at Sheridan College for providing financial support to hire student researchers who contributed to data annotation activities.
\subsection{Ethical Approval}
Not applicable.
\subsection{Informed Consent}
Not applicable.

\subsection{Author Contributions}

Kobra Khanmohammadi: Conceptualization, methodology, data analysis, supervision, writing, reviewing and editing.

Pooria Roy: Data curation, methodology, implementation, formal analysis, writing.

Raphaël Khoury: Supervision, conceptualization, methodology, writing, reviewing and editing.

Abdelwahab Hamou-Lhadj: Supervision, validation, reviewing and editing.

Wilfried Patrick Konan: Data analysis, implementation.

Alexander Da Re: Data annotation, validation.

Nicholas Rebelo Melo: Data annotation, validation.
\subsection{Data Availability Statement}
The dataset and codes in this study are publicly available on the authors’ GitHub repository at \url{https://github.com/regularpooria/wildcode}, as referenced in the paper.

\subsection{Conflict of Interest}

The authors declare that they have no conflict of interest.


\bibliographystyle{splncs04}
\bibliography{bibliography}

@inproceedings{original,
  title={WildCode: An Empirical Analysis of Code Generated by ChatGPT},
  author={Kobra Khanmohammadi and Pooria Roy and  Raphael Khoury and Abdelwahab Hamou-Lhadj and Wilfried Patrick Konan},
  booktitle={Proceedings of the 18th International Symposium on Foundations and Practice of Security (FPS - 2025)},
  pages={},
  year={2025}
}

@article{kruskal1952use,
  title={Use of ranks in one-criterion variance analysis},
  author={Kruskal, William H. and Wallis, W. Allen},
  journal={Journal of the American Statistical Association},
  volume={47},
  number={260},
  pages={583--621},
  year={1952},
  publisher={Taylor \& Francis}
}

@article{kullback1951information,
  title={On information and sufficiency},
  author={Kullback, Solomon and Leibler, Richard A},
  journal={The Annals of Mathematical Statistics},
  volume={22},
  number={1},
  pages={79--86},
  year={1951}
}

@inproceedings{li2024cosec,
  title={CoSec: On-the-Fly Security Hardening of Code LLMs via Supervised Co-decoding},
  author={Li, Dong and Yan, Meng and Zhang, Yaosheng and Liu, Zhongxin and Liu, Chao and Zhang, Xiaohong and Chen, Ting and Lo, David},
  booktitle={Proceedings of the 33rd ACM SIGSOFT International Symposium on Software Testing and Analysis},
  pages={1428--1439},
  year={2024}
}

@article{bai2024apilot,
  title={APILOT: Navigating Large Language Models to Generate Secure Code by Sidestepping Outdated API Pitfalls},
  author={Bai, Weiheng and Xuan, Keyang and Huang, Pengxiang and Wu, Qiushi and Wen, Jianing and Wu, Jingjing and Lu, Kangjie},
  journal={arXiv preprint arXiv:2409.16526},
  year={2024}
}

@article{zeng2025inducing,
  title={Inducing Vulnerable Code Generation in LLM Coding Assistants},
  author={Zeng, Binqi and Zhang, Quan and Zhou, Chijin and Go, Gwihwan and Jiang, Yu and Shi, Heyuan},
  journal={arXiv preprint arXiv:2504.15867},
  year={2025}
}

@misc{owasp-redos,
  author       = {{OWASP Foundation}},
  title        = {Regular Expression Denial of Service (ReDoS)},
  year         = {2025},
  howpublished = {{\url{owasp.org/www-community/attacks/Regular\_expression\_Denial\_of\_Service\_-\_ReDoS}}},
  note         = {Accessed: 2025-08-04}
}

@article{zhao2024wildchat,
  title={Wildchat: 1m chatgpt interaction logs in the wild},
  author={Zhao, Wenting and Ren, Xiang and Hessel, Jack and Cardie, Claire and Choi, Yejin and Deng, Yuntian},
  journal={arXiv preprint arXiv:2405.01470},
  year={2024}
}

@article{sajadi2025llms,
  title={Do LLMs consider security? an empirical study on responses to programming questions},
  author={Sajadi, Amirali and Le, Binh and Nguyen, Anh and Damevski, Kostadin and Chatterjee, Preetha},
  journal={Empirical Software Engineering},
  volume={30},
  number={3},
  pages={101},
  year={2025},
  publisher={Springer}
}

@inproceedings{spracklen2025we,
  title={We Have a Package for You! A Comprehensive Analysis of Package Hallucinations by Code Generating {LLMs}},
  author={Spracklen, Joseph and Wijewickrama, Raveen and Sakib, AHM Nazmus and Maiti, Anindya and Viswanath, Bimal},
  booktitle={34th USENIX Security Symposium},
  pages={3687--3706},
  year={2025}
}

@inproceedings{khoury2023secure,
  title={How secure is code generated by chatgpt?},
  author={Khoury, Rapha{\"e}l and Avila, Anderson R and Brunelle, Jacob and Camara, Baba Mamadou},
  booktitle={2023 IEEE international conference on systems, man, and cybernetics (SMC)},
  pages={2445--2451},
  year={2023},
  organization={IEEE}
}

@article{tihanyi2023new,
  title={A new era in software security: Towards self-healing software via large language models and formal verification},
  author={Tihanyi, Norbert and Jain, Ridhi and Charalambous, Yiannis and Ferrag, Mohamed Amine and Sun, Youcheng and Cordeiro, Lucas C},
  journal={arXiv preprint arXiv:2305.14752},
  year={2023}
}

@article{fu2025security,
  title={Security Weaknesses of Copilot-Generated Code in GitHub Projects: An Empirical Study},
  author={Fu, Yujia and Liang, Peng and Li, Zengyang and Shahin, Mojtaba and Yu, Jiaxin and Chen, Jinfu},
  journal={ACM Transactions on Software Engineering and Methodology},
  year={2025},
  publisher={ACM New York, NY}
}

@misc{github2025survey,
  author       = {{GitHub}},
  title        = {Survey reveals AI’s impact on the developer experience},
  year         = {2025},
  howpublished = {\url{https://github.blog/news-insights/research/survey-reveals-ais-impact-on-the-developer-experience/}},
  note         = {Accessed: 2025-05-26}
}

@misc{aws2024codewhisperer,
  author       = {{AWS DevOps Blog}},
  title        = {Introducing Amazon CodeWhisperer Dashboard and CloudWatch Metrics},
  year         = {2024},
  howpublished = {\url{https://aws.amazon.com/blogs/devops/introducing-amazon-codewhisperer-dashboard-and-cloudwatch-metrics/}}
}

@article{yeticstiren2023evaluating,
  title={Evaluating the code quality of ai-assisted code generation tools: An empirical study on github copilot, amazon codewhisperer, and chatgpt},
  author={Yeti{\c{s}}tiren, Burak and {\"O}zsoy, I{\c{s}}{\i}k and Ayerdem, Miray and T{\"u}z{\"u}n, Eray},
  journal={arXiv preprint arXiv:2304.10778},
  year={2023}
}

@inproceedings{he2023large,
  title={Large language models for code: Security hardening and adversarial testing},
  author={He, Jingxuan and Vechev, Martin},
  booktitle={2023 ACM SIGSAC Conference on Computer and Communications Security},
  pages={1865--1879},
  year={2023}
}

@inproceedings{nazzal2024promsec,
  title={PromSec: Prompt Optimization for Secure Generation of Functional Source Code with Large Language Models ({LLM}s)},
  author={Nazzal, Mahmoud and Khalil, Issa and Khreishah, Abdallah and Phan, NhatHai},
  booktitle={2024 on ACM SIGSAC Conference on Computer and Communications Security},
  pages={2266--2280},
  year={2024}
}

@article{ashrafi2025enhancing,
  title={Enhancing LLM Code Generation: A Systematic Evaluation of Multi-Agent Collaboration and Runtime Debugging for Improved Accuracy, Reliability, and Latency},
  author={Ashrafi, Nazmus and Bouktif, Salah and Mediani, Mohammed},
  journal={arXiv preprint arXiv:2505.02133},
  year={2025}
}

@article{xu2024prosec,
  title={ProSec: Fortifying Code LLMs with Proactive Security Alignment},
  author={Xu, Xiangzhe and Su, Zian and Guo, Jinyao and Zhang, Kaiyuan and Wang, Zhenting and Zhang, Xiangyu},
  journal={arXiv preprint arXiv:2411.12882},
  year={2024}
}

@article{fakhoury2024llm,
  title={Llm-based test-driven interactive code generation: User study and empirical evaluation},
  author={Fakhoury, Sarah and Naik, Aaditya and Sakkas, Georgios and Chakraborty, Saikat and Lahiri, Shuvendu K},
  journal={IEEE Transactions on Software Engineering},
  year={2024},
  publisher={IEEE}
}

@article{huang2024bias,
  title={Bias testing and mitigation in llm-based code generation},
  author={Huang, Dong and Zhang, Jie M and Bu, Qingwen and Xie, Xiaofei and Chen, Junjie and Cui, Heming},
  journal={ACM Transactions on Software Engineering and Methodology},
  year={2024},
  publisher={ACM New York, NY}
}

@article{liu2023your,
  title={Is your code generated by chatgpt really correct? rigorous evaluation of large language models for code generation},
  author={Liu, Jiawei and Xia, Chunqiu Steven and Wang, Yuyao and Zhang, Lingming},
  journal={Advances in Neural Information Processing Systems},
  volume={36},
  pages={21558--21572},
  year={2023}
}

@misc{SafeRegex,
  author={Joe Kutner},
  title = {SafeRegex},
  howpublished ={https://github.com/jkutner/saferegex},
  year = {2018}
}

@inproceedings{Recue,
author = {Shen, Yuju and Jiang, Yanyan and Xu, Chang and Yu, Ping and Ma, Xiaoxing and Lu, Jian},
title = {ReScue: crafting regular expression DoS attacks},
year = {2018},
isbn = {9781450359375},
publisher = {Association for Computing Machinery},
address = {New York, NY, USA},
booktitle = {Proceedings of the 33rd ACM/IEEE International Conference on Automated Software Engineering},
pages = {225–235},
numpages = {11},
location = {Montpellier, France},
series = {ASE '18}
}

@inproceedings{redoshunter,
  title={{ReDoSHunter}: A combined static and dynamic approach for regular expression {DoS} detection},
  author={Li, Yeting and Chen, Zixuan and Cao, Jialun and Xu, Zhiwu and Peng, Qiancheng and Chen, Haiming and Chen, Liyuan and Cheung, Shing-Chi},
  booktitle={30th USENIX Security Symposium},
  pages={3847--3864},
  year={2021}
}

@inproceedings{revealer,
  title={Revealer: Detecting and exploiting regular expression denial-of-service vulnerabilities},
  author={Liu, Yinxi and Zhang, Mingxue and Meng, Wei},
  booktitle={2021 IEEE Symposium on Security and Privacy (SP)},
  pages={1468--1484},
  year={2021},
  organization={IEEE}
}

@article{log4shell,
  title={The race to the vulnerable: Measuring the log4j shell incident},
  author={Hiesgen, Raphael and Nawrocki, Marcin and Schmidt, Thomas C and W{\"a}hlisch, Matthias},
  journal={arXiv preprint arXiv:2205.02544},
  year={2022}
}

@article{sayar2023depth,
  title={An in-depth study of java deserialization remote-code execution exploits and vulnerabilities},
  author={Sayar, Imen and Bartel, Alexandre and Bodden, Eric and Le Traon, Yves},
  journal={ACM Transactions on Software Engineering and Methodology},
  volume={32},
  number={1},
  pages={1--45},
  year={2023},
  publisher={ACM New York, NY}
}

@article{Etsenake2024,
  title={Understanding the Human-LLM Dynamic: A Literature Survey of LLM Use in Programming Tasks},
  author={Deborah Etsenake and Meiyappan Nagappan},
  journal={ArXiv},
  year={2024},
  volume={abs/2410.01026},
  url={https://api.semanticscholar.org/CorpusID:273026291}
}

@inproceedings{peng2025cweval,
  title     = {CWEval: Outcome-Driven Evaluation on Functionality and Security of LLM Code Generation},
  author    = {Peng, Jinjun and others},
  booktitle = {2025 IEEE/ACM International Workshop on Large Language Models for Code (LLM4Code)},
  year      = {2025},
  publisher = {IEEE}
}

@article{toth2024php,
  title   = {LLMs in Web Development: Evaluating LLM-Generated PHP Code Unveiling Vulnerabilities and Limitations},
  author  = {T{\'o}th, G{\'a}bor and Bisztray, D{\'a}vid and Erdodi, L{\'a}szl{\'o}},
  journal = {arXiv preprint arXiv:2404.14674},
  year    = {2024}
}

@inproceedings{dora2025hidden,
  title     = {The Hidden Risks of LLM-Generated Web Application Code: A Security-Centric Evaluation},
  author    = {Dora, Swaroop and others},
  booktitle = {International Conference on Information Systems Security},
  year      = {2025},
  publisher = {Springer}
}

@article{kharma2025securityquality,
  title   = {Security and Quality in LLM-Generated Code: A Multi-Language, Multi-Model Analysis},
  author  = {Kharma, Mohammed and others},
  journal = {arXiv preprint arXiv:2502.01853},
  year    = {2025}
}




\end{document}